\documentclass[twocolumn]{aastex63}
\usepackage{amsmath}
\usepackage{graphicx}

\received{*}
\revised{*}
\accepted{*}
\submitjournal{ApJ}
\shorttitle{Radio echo in the turbulent corona}
\shortauthors{Kuznetsov et al.}

\begin{document}
\title{Radio Echo in the Turbulent Corona and Simulations of Solar Drift-Pair Radio Bursts}

\correspondingauthor{Alexey A. Kuznetsov}
\email{a\_kuzn@iszf.irk.ru}

\author[0000-0001-8644-8372]{Alexey A. Kuznetsov}
\affiliation{Institute of Solar-Terrestrial Physics, Irkutsk 664033, Russia}
\affiliation{School of Physics \& Astronomy, University of Glasgow, Glasgow, G12 8QQ, UK}

\author[0000-0002-4389-5540]{Nicolina Chrysaphi}
\affiliation{School of Physics \& Astronomy, University of Glasgow, Glasgow, G12 8QQ, UK}

\author[0000-0002-8078-0902]{Eduard P. Kontar}
\affiliation{School of Physics \& Astronomy, University of Glasgow, Glasgow, G12 8QQ, UK}

\author[0000-0001-7856-084X]{Galina Motorina}
\affiliation{Astronomical Institute of the Czech Academy of Sciences, 251 65 Ondrejov, Czech Republic}
\affiliation{Central Astronomical Observatory at Pulkovo of Russian Academy of Sciences, St. Petersburg, 196140, Russia}

\begin{abstract}
Drift-pair bursts are an unusual type of solar low-frequency radio emission, which appear in the dynamic spectra as two parallel drifting bright stripes separated in time. Recent imaging spectroscopy observations allowed for the quantitative characterization of the drifting pairs in terms of source size, position, and evolution. Here, the drift-pair parameters are qualitatively analyzed and compared with the newly-developed Monte Carlo ray-tracing technique simulating radio-wave propagation in the inhomogeneous anisotropic turbulent solar corona. The results suggest that the drift-pair bursts can be formed due to a combination of the refraction and scattering processes, with the trailing component being the result of turbulent reflection (turbulent radio echo). The formation of drift-pair bursts requires an anisotropic scattering with the level of plasma density fluctuations comparable to that in type III bursts, but with a stronger anisotropy at the inner turbulence scale. The anisotropic radio-wave scattering model can quantitatively reproduce the key properties of drift-pair bursts: the apparent source size and its increase with time at a given frequency, the parallel motion of the source centroid positions, and the delay between the burst components. The trailing component is found to be virtually co-spatial and following the main component. The simulations suggest that the drift-pair bursts are likely to be observed closer to the disk center and below 100 MHz due to the effects of free-free absorption and scattering. The exciter of drift-pairs is consistent with propagating packets of whistlers, allowing for a fascinating way to diagnose the plasma turbulence and the radio emission mechanism.
\end{abstract}


\section{Introduction}
Radio bursts occur commonly in the outer solar corona following strong solar flares and even during periods of weak solar activity. At the same time, radio emission produced close to plasma frequency experiences strong refraction and scattering during propagation through the inhomogeneous turbulent corona, which can lead to a significant (or even dominant) effect on the observed positions and sizes of the radio sources, as well as on the observed time profiles of the radio bursts.

Drift-pair bursts are a rare and mysterious type of fine spectral structures in the low-frequency domain of solar radio emissions. First identified by \citet{roberts_1958} spectrally, they appear in the dynamic spectrum as two parallel frequency-drifting bright stripes separated in time, where the trailing stripe seems to repeat the morphology of the leading one with a typical delay of $\sim 1-2$ s. Until recently, the drift-pair bursts have mainly been studied using dynamic radio spectroscopy, but with limited or no imaging information. A breakthrough was achieved due to the imaging spectroscopy observations with the LOw-Frequency ARray \cite[LOFAR,][]{vanHaarlem_2013}, which enabled \citet{kuznetsov_2019} to resolve, or the first time, the evolution of the radio sources both at a fixed frequency and along the drifting pair components. These observations also indicated the importance of radio-scattering effects, including anisotropic scattering.

Recently, \citet{kontar_2019} have developed a theoretical framework and computer codes simulating the radio-wave transport by accounting for anisotropic scattering as well as large-scale refraction,and used the model to investigate the observed characteristics of the solar type III radio bursts. In this work, we apply this model to the drift-pair bursts, with an aim to reproduce their distinctive double-pulse structure. In Section \ref{observations}, we briefly summarize the observed properties of these bursts. In Section \ref{model}, we describe the simulation approach and the model assumptions. In Section \ref{results}, we present the simulation results and compare them to the observations. In Section \ref{discussion}, we discuss the obtained results and their implications for understanding the origin of the drift-pair bursts.

\section{Observed properties of the drift-pair bursts}\label{observations}
Basic characteristics of the drift-pair bursts are sum\-ma\-rized, e.g., in the papers of \citet{melrose_1982}, \citet{melnik_2005}, and \citet{kuznetsov_2019}. A typical drifting pair consists of two short narrow-band frequency-drifting stripes; both positive (which are more common) and negative frequency drifts are observed. The bursts occur in the frequency range of $\sim 10-100$ MHz. The frequency drift rates tend to increase with the emission frequency, being generally of about $1-2$ MHz $\textrm{s}^{-1}$ at the frequency of $\sim 30$ MHz, which is about ten times higher than typical drift rates of type II bursts and three times lower than typical drift rates of type III bursts at the same frequencies \citep{mclean_1985}. The circular polarization degree varies from $\sim 10$\% to $\sim 50$\%, which favors the fundamental plasma emission mechanism \citep{suzuki_1979, dulk_1984}.

The most intriguing feature of the drift-pair bursts is their double structure, where the second (trailing) component looks like an almost exact repetition of the first (leading) one, with the same start and end frequencies, and the same frequency drift rate. The duration of each component at a fixed frequency is about 1 s, and the delay between the components is about $1-2$ s at all frequencies (although \citealt{melnik_2005} reported a weak decrease of this delay with frequency). Interestingly, the bursts with negative frequency drift are usually shorter and more narrowband than the bursts with positive frequency drift.

\citet{suzuki_1979}, using imaging observations with the Culgoora radioheliograph, discovered that the emission sources of the first and second components of a drifting pair (at the same frequency) coincide spatially. Recently, \citet{kuznetsov_2019} analyzed the multi-frequency imaging observations of drift-pair bursts with LOFAR and found that the sources of both components of a drifting pair propagate (with a certain delay) in the same direction along the same trajectory.

\begin{figure*}
\centerline{\includegraphics[width=0.49\linewidth]{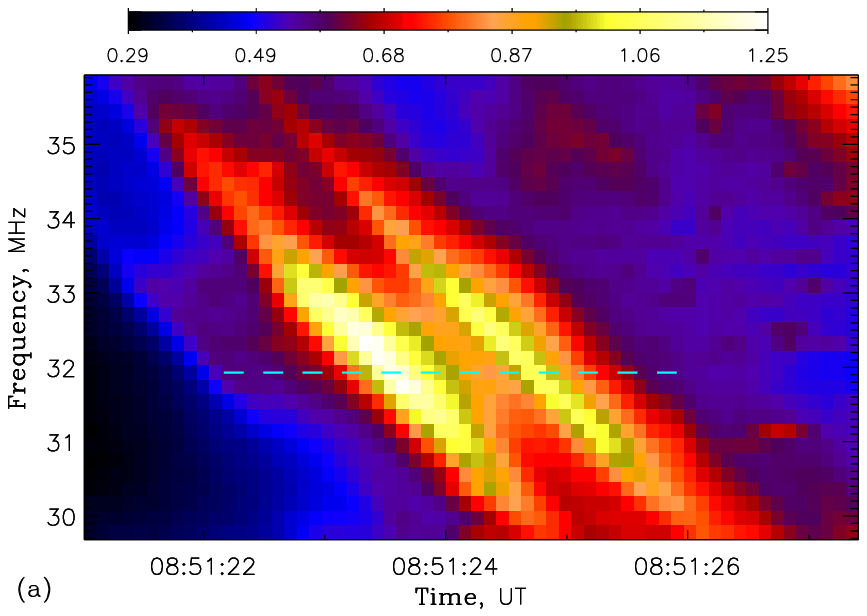}~
\includegraphics[width=0.49\linewidth]{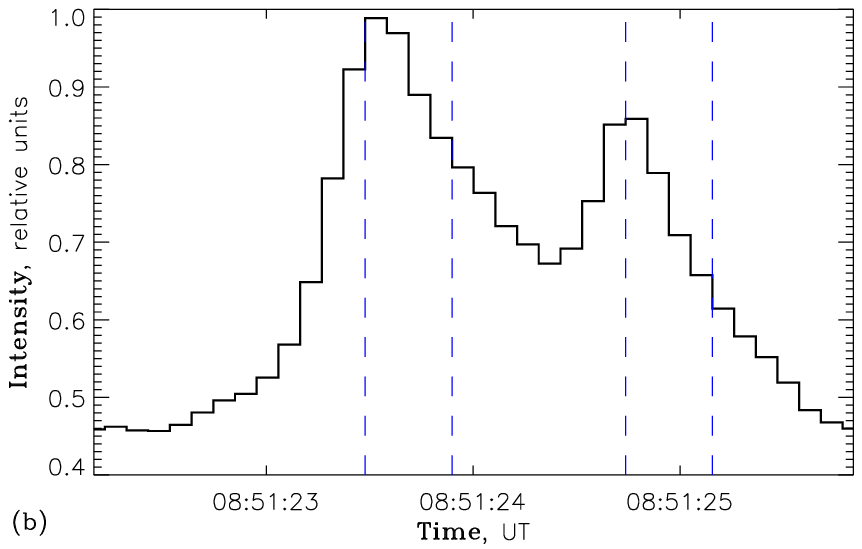}}
\centerline{\includegraphics[width=0.49\linewidth]{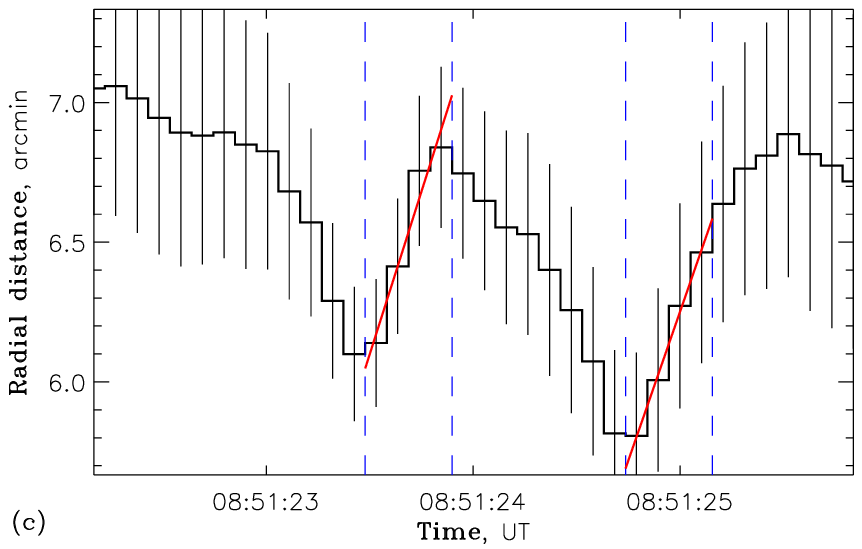}~
\includegraphics[width=0.49\linewidth]{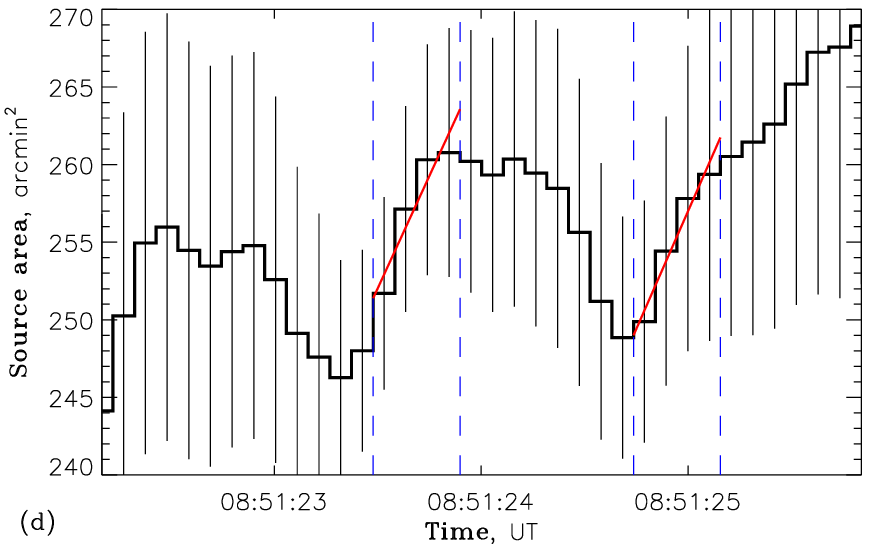}}
\caption{a) Dynamic spectrum of solar radio emission with a drift-pair burst recorded with LOFAR on 12 July 2017 (Sun-integrated, in relative units). b) Time profile of the radio flux at a single frequency (32 MHz). c) Corresponding time profile of the visible radio source position (distance from the solar disk center). d) Corresponding time profile of the visible radio source area (at half-maximum level). Red lines represent linear fits to the time profiles of the source parameters in the shown intervals. The error bars represent one standard deviation.}
\label{obs_prof}
\end{figure*}

Figure \ref{obs_prof} presents the dynamic spectrum of a typical drift-pair burst observed with LOFAR (see \citealt{kuznetsov_2019} for the description of the instrument configuration and the observed event) as well as the time evolution of the radio flux and the emission source parameters (position and size) at a fixed frequency. The source parameters were determined by fitting the spatially-resolved LOFAR data with an elliptical Gaussian \citep{kontar_2017}. The delay between the burst components is about $1.2$ s. A notable feature is that at the decay phases of both components (shortly after the intensity peaks), the emission source demonstrates a clear radial (outward) motion and an increase in size; the source speed and area expansion rate are about $\mathrm{d}r/\mathrm{d}t\simeq 2.2$ arcmin $\textrm{s}^{-1}\simeq c/3$ (where $c$ is the speed of light) and $\mathrm{d}A/\mathrm{d}t\simeq 30$ $\textrm{arcmin}^2$ $\textrm{s}^{-1}$, respectively. This behaviour is reminiscent of the source dynamics detected in type IIIb bursts \citep{kontar_2017, sharykin_2018} and can be explained naturally by scattering of radio emission during propagation. The background continuum emission has a source position and size not very different from those of the drift-pair bursts \citep[cf.][]{suzuki_1979}, and probably has a similar physical origin, but without the fine structure. Unfortunately, there are no LOFAR polarization data for this event.

Following \citet{sharykin_2018}, we estimate the anisotropy of the scattering process. The measured area of the emission source $A_{\mathrm{vis}}$ (see Figure \ref{obs_prof}d) is related to its actual area $A_{\mathrm{real}}$ as $A_{\mathrm{vis}}\simeq A_{\mathrm{real}}+A_{\mathrm{beam}}$, where $A_{\mathrm{beam}}$ is the instrument beam area. For the visible source area $A_{\mathrm{vis}}\simeq 250$ $\textrm{arcmin}^2$ and the LOFAR beam area $A_{\mathrm{beam}}\simeq 100$ $\textrm{arcmin}^2$, we obtain $A_{\mathrm{real}}\simeq 150$ $\textrm{arcmin}^2$, which corresponds (for a nearly circular source) to the linear source size of about 14 arcmin.  This estimation includes the source expansion due to scattering and represents the source size across the line-of-sight $\Delta r_{\bot}$. On the other hand, the source extent along the line-of-sight $\Delta r_{\mathrm{LOS}}$ (including the effects of scattering) cannot exceed $c\Delta t$, where $\Delta t$ is the burst (one component) duration. For $\Delta t\simeq 0.6$ s (see Figure \ref{obs_prof}b), we obtain $\Delta r_{\mathrm{LOS}}\lesssim 4$ arcmin. Since $\Delta r_{\bot}\gg\Delta r_{\mathrm{LOS}}$, the radio-wave scattering in the corona should be highly anisotropic: scattering perpendicular to the line-of-sight is much more efficient than that in the parallel to the line-of-sight direction.

\begin{figure}
\centerline{\includegraphics{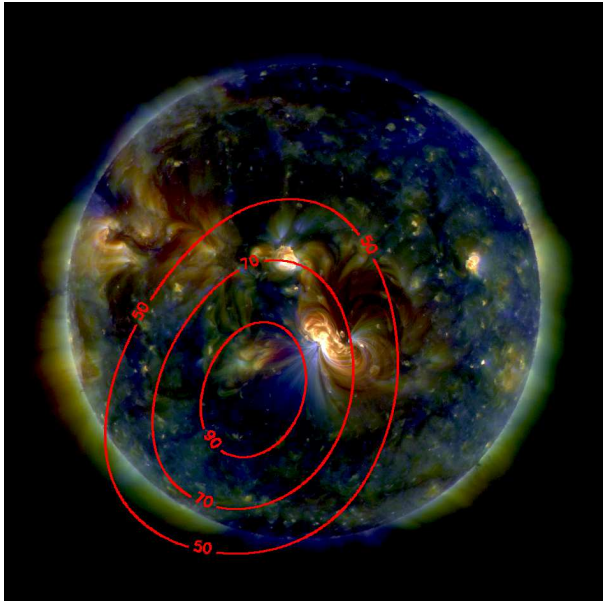}}
\caption{LOFAR radio map (intensity contours drawn at the levels of 50, 70, and 90\% of the maximum intensity) overplotted on the combined SDO/AIA EUV image. The radio map corresponds to the frequency of 32 MHz and time of 08:51:23.55 UT (the first peak of the burst shown in Figure \protect\ref{obs_prof}). The background EUV map includes the SDO/AIA images at 211 {\AA} (red), 193 {\AA} (green), and 171 {\AA} (blue).}
\label{obs_img}
\end{figure}

Figure \ref{obs_img} presents the LOFAR image (intensity contours) overplotted on the combined three-wavelength SDO/AIA EUV image \citep{lemen_2012}. The radio map corresponds to the frequency of 32 MHz and the first peak of the burst shown in Figure \ref{obs_prof}. At the metric wavelengths, the apparent radio source position can be affected significantly by refraction in the ionosphere;  in the considered event, the resulting source shift could be up to $7-8$ arcmin \citep[cf.][]{gordovskyy_2019}\footnote{ The radio image in Figure \ref{obs_img} was not corrected for possible ionospheric effects.}. In any case, the source centroid seems to be located not far from the solar disk center (is projected on the solar disk) and  the emission source is likely to be associated with the large active region. We also highlight that ionospheric refraction affects the source positions of the first and second component of a drift-pair burst (at the same frequency) in the same way; i.e., it does not affect the relative positions of the component sources.

Initially, drift-pair bursts were interpreted by reflection at the plasma level \citep[radio echo effect;][]{roberts_1958}, so that the second component represents a signal reflected from the lower (denser) layers of the solar corona, where the local plasma frequency is equal to the radio-wave frequency. However, this model was later questioned \citep[e.g.,][]{melrose_1982} because, on the first glance, different ray paths should result in different visible source positions of the first (direct) and second (reflected) burst components, especially for  the harmonic emission and sources located far from the solar disk center. In addition, the reflected signal should be considerably weakened and smoothed in comparison with the direct one. However, accounting for the scattering effects allows for a more realistic model where both the direct and reflected radio waves (as well as the waves propagating downwards before reflection) are scattered. In such scenario, the apparent source positions and light curves can be strongly altered by scattering; e.g., the visible sources of both burst components are expected to be located at the distance of the ``last-scattering surface'' \citep{chrysaphi_2018, kontar_2019}. We now explore whether the radio echo with anisotropic turbulence can reproduce the observed characteristics of the drift-pair bursts.

\section{Monte Carlo simulation set-up}\label{model}
To study the radio-wave propagation in the turbulent plasma of the solar corona, we employ the 3D Monte Carlo ray-tracing technique, as presented by \citet{kontar_2019}. Namely, we numerically integrate the Langevin equations describing the time evolution of the radio-wave packets (``photons'') both in the coordinate space and in the space of wave vectors. The numerical code includes effects like: a) refraction due to large-scale gradual variation of the plasma density, b) systematic change of the wave vector due to elastic scattering, c) random changes of the propagation direction due to scattering, and d) collisional damping.  For simplicity (and also because we currently do not have spatially resolved polarization observations), we consider the intensity transfer only, like in an unmagnetized plasma. The same code was successfully used by \citet{kontar_2019} to reproduce the characteristics of solar type III burst sources.

We consider an anisotropic (but axially symmetric) plasma turbulence described by the spectrum of density fluctuations in the form
\begin{equation}\label{anisoS}
S(\boldsymbol{q})=S\left[\left(q_{\perp}^2+\alpha^{-2}q_{\parallel}^2\right)^{1/2}\right],
\end{equation}
where $\boldsymbol{q}$ is the wave vector of the density fluctuations, $q_{\perp}$ and $q_{\parallel}$ are its components in the perpendicular and parallel (to the magnetic field) directions, respectively, and $\alpha$ is the anisotropy parameter. As demonstrated by \citet{kontar_2019}, preferable scattering in the perpendicular direction \citep[which agrees nicely not only with the above estimations for the drift-pair burst source in Section \ref{observations} but also with the radio imaging of compact sources via the corona by][]{hewish_1958, baselyan_1971, anantharamaiah_1994} requires $\alpha\ll 1$, because in this case the perpendicular plasma fluctuations dominate, too. In turn, following the radio observations \citep[e.g.,][]{woo_1979}, the dependence of the density fluctuation spectrum (\ref{anisoS}) on the ``effective'' wave number $S(\tilde q)$ is taken to be a power-law Kolmogorov spectrum in the range of wavelengths from $l_{\mathrm{i}}$ to $l_{\mathrm{o}}$ (see below). Then the angular scattering rate will be proportional to the spectrum-averaged mean wavenumber \citep[see Appendices in the paper of][]{kontar_2019}
\begin{equation}
\overline{q\epsilon^2}=\int q S(\boldsymbol{q})\frac{\mathrm{d}^3\boldsymbol{q}}{(2\pi)^3},
\end{equation}
where the level of density fluctuations is characterized by the parameter $\epsilon$:
\begin{equation}
\epsilon^2=\frac{\left<\delta n^2\right>}{\left<n\right>^2}=\int S(\boldsymbol{q})\frac{\mathrm{d}^3\boldsymbol{q}}{(2\pi)^3},
\end{equation}
and $n$ is the electron plasma density. Noteworthy, in some earlier works \citep[e.g.,][]{steinberg_1971, chrysaphi_2018}, the scattering rate is also characterised by $\epsilon^2/h$ with a Gaussian spectrum of density fluctuations, where $h$ is the characteristic correlation length of the fluctuations.

In this work, we consider a spherically symmetric solar corona with a radial magnetic field; i.e., the anisotropic turbulence (\ref{anisoS}) is always aligned with respect to the local radial direction. The plasma density decreases with height according to the density model by \citet{parker_1960} with refinements by \citet{mann_1999}, which was approximated by an analytical model \citep[see Equation (43) in the paper of ][]{kontar_2019}. Both the inner ($l_{\mathrm{i}}$) and outer ($l_{\mathrm{o}}$) scales of the plasma density fluctuations increase with height following the empirical relations by \citet{manoharan_1987, coles_1989, wohlmuth_2001}; e.g., for the heliocentric distances from 1 to 100 $R_{\sun}$, the inner scale $l_{\mathrm{i}}$ (which is the primary parameter determining the scattering rate) increases linearly from 1 to 100 km, and at the same time, the level of density fluctuations $\epsilon$ and anisotropy $\alpha$ are assumed to be the same at all radial distances.

We consider the levels of density fluctuations $\epsilon=0$ (i.e., no fluctuations and hence no scattering) and $\epsilon=0.8$; The level of density fluctuations $\epsilon=0.8$ might appear rather high; however, we note that this value is dependent on the adopted turbulence scale, because the radio scattering observations allow us to estimate the parameter $\overline{q\epsilon^2}\sim\epsilon^2/h$ only. Therefore, using $\epsilon=0.8$ together with the above-mentioned turbulence model by \citet{manoharan_1987, coles_1989, wohlmuth_2001} is equivalent to using, e.g., a density fluctuations level that is 10 times lower (i.e., $\epsilon=0.08$) together with 100 times shorter fluctuation wavelengths.

 Although we do not consider scattering of emission on ``fibrous'' structures of the plasma density such as streamers or overdense magnetic loops \citep{riddle_1974, bougeret_1977, robinson_1983}, any density fluctuation model with an appropriate scattering rate $\overline{q\epsilon^2}$ and anisotropy would produce similar results. Propagation of radio emission through a ``fibrous medium'' comprised of multiple quasi-randomly distributed magnetic tubes could provide scattering qualitatively similar to the effect of anisotropic (field-aligned) plasma turbulence \citep{riddle_1974, bougeret_1977, robinson_1983}. However, since these long-living structures have characteristic scales (hundreds to thousands of km) much larger than those of the irregular turbulence, their effect is expected to be much weaker. To provide the scattering rate (i.e., the $\overline{q\epsilon^2}$ parameter) comparable with that of the observations, magnetic tubes would need to have the density contrast of $\delta n/n\gg 1$ (e.g., \citealt{robinson_1983} considered a 25-fold increase of the plasma density over dense fibres). Existence of such structures in the solar corona is not supported by EUV observations \citep[e.g.,][]{motorina_2020}.

We start the simulations with a point source located at a certain heliocentric distance $r_0$ and  heliocentric longitude $\theta_0$. We consider a number ($\sim 10^4$) of photons with the same frequency (which is not changed during propagation); they are injected at the source point simultaneously (i.e., initially the pulse has a delta-function time profile) and initially have an isotropic distribution in the wave vector. The source location is chosen to provide a certain ($>1$) ratio of the emission frequency $f$ to the electron plasma frequency $f_{\mathrm{pe}}(r_0)$; the photons that initially propagate downwards can reach the observer only due to refraction and/or scattering. In each simulation run, the photons are traced until the scattering becomes insignificant, and then the resulting light curve and apparent radio brightness map for an observer at the Earth are reconstructed. Free-free absorption of the photons due to plasma collisions is included as the weight of the photons \citep[see, e.g.,][]{jeffrey_2011, kontar_2019}; the plasma temperature which affects the collisional damping is assumed to be $T=1$ MK. The initial ratio of the emission frequency to the local plasma frequency is assumed to be $f/f_{\mathrm{pe}}(r_0)=1.05-1.10$, which means that fundamental plasma emission is considered.

\begin{figure*}
\centerline{\includegraphics[width=0.3\linewidth]{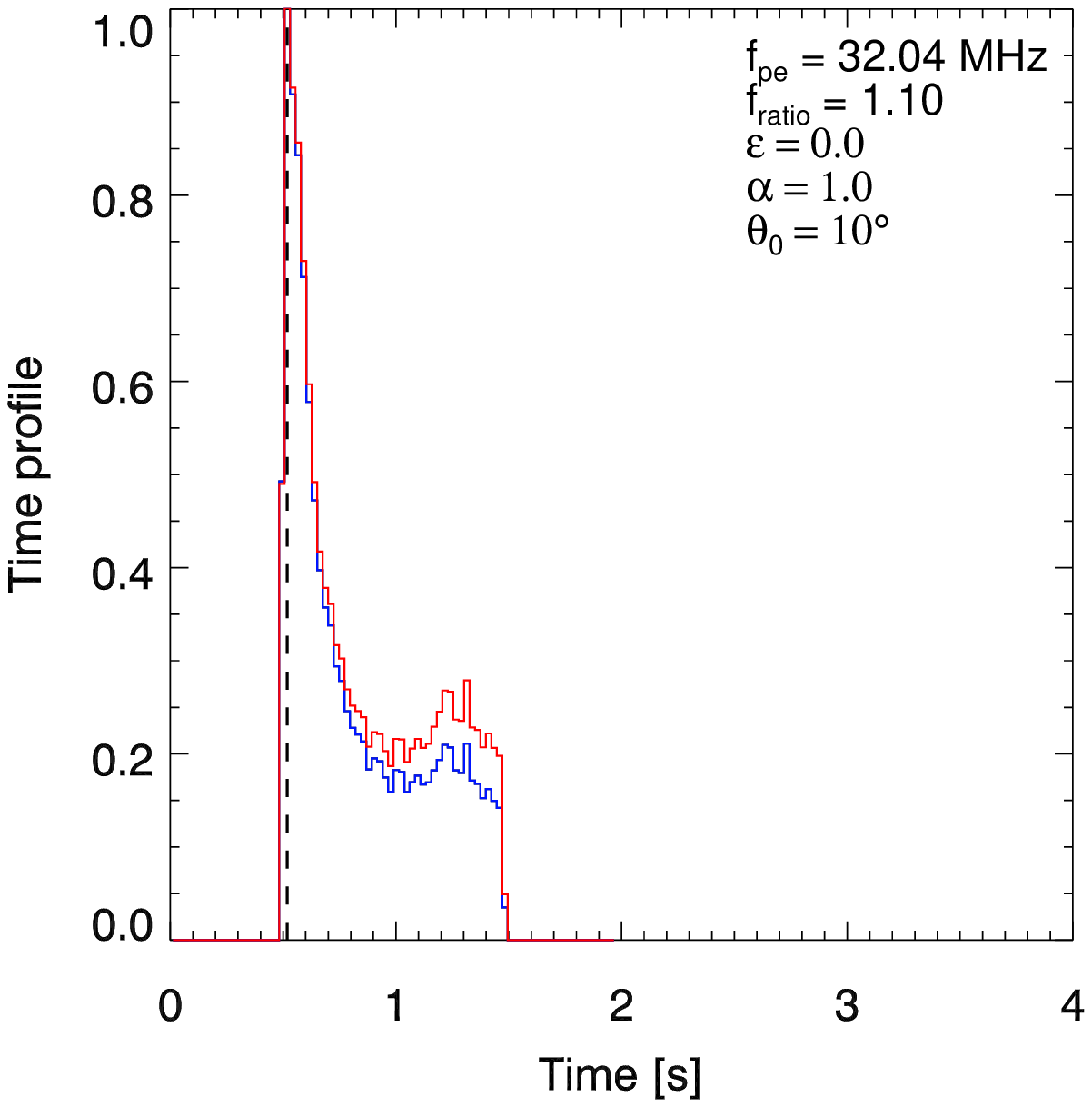}~
\includegraphics[width=0.3\linewidth]{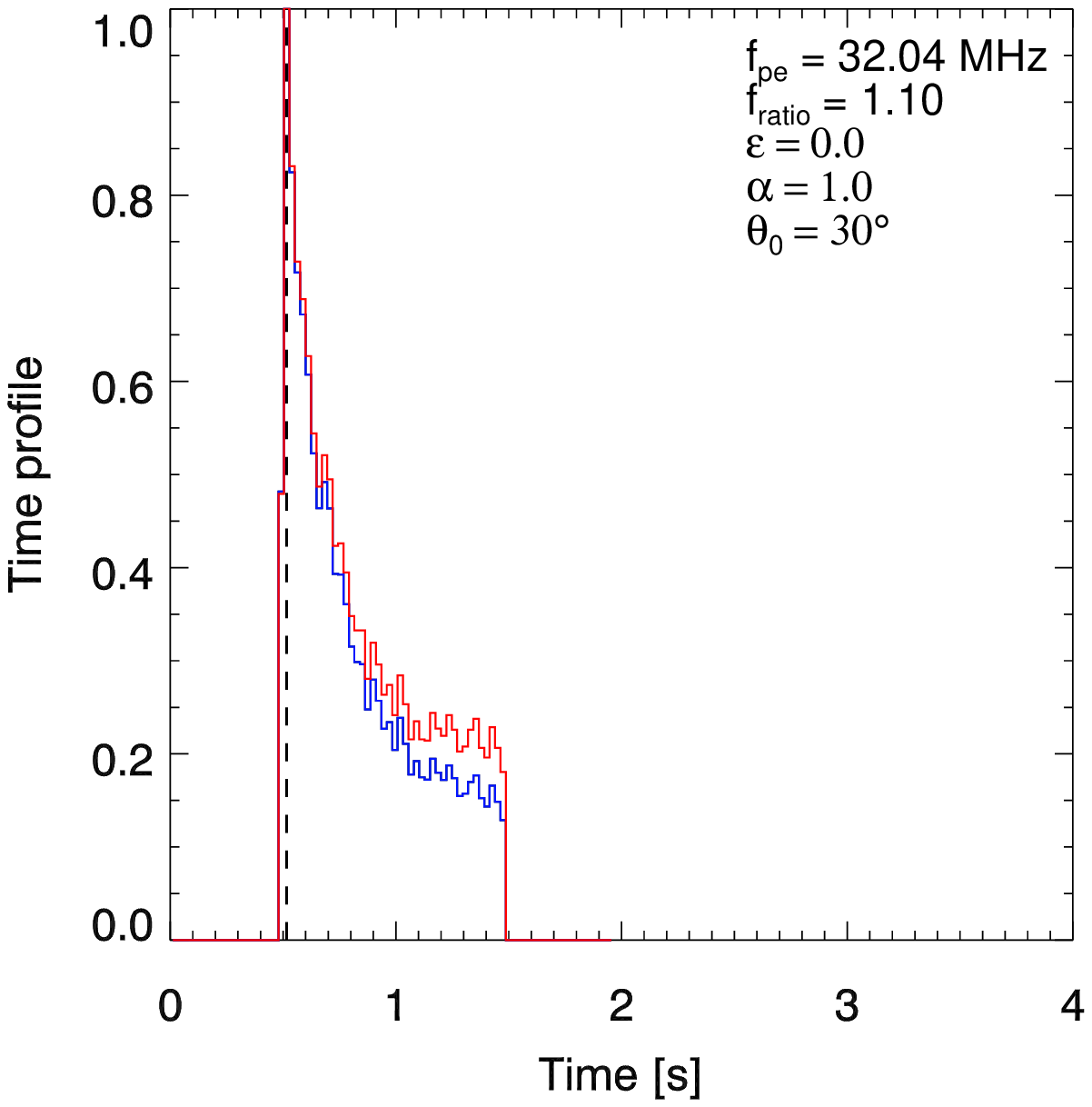}}
\centerline{\includegraphics[width=0.3\linewidth]{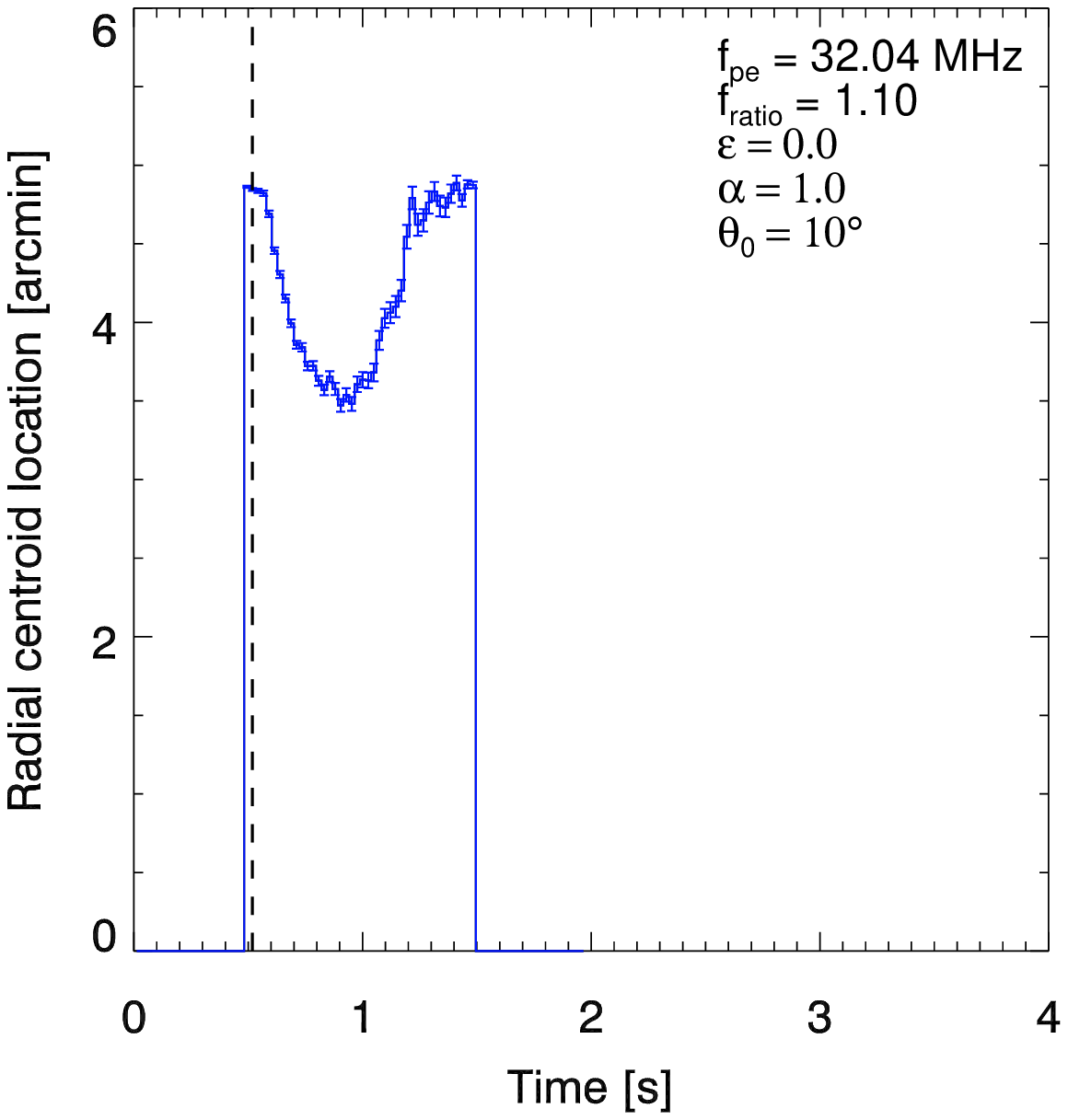}~
\includegraphics[width=0.3\linewidth]{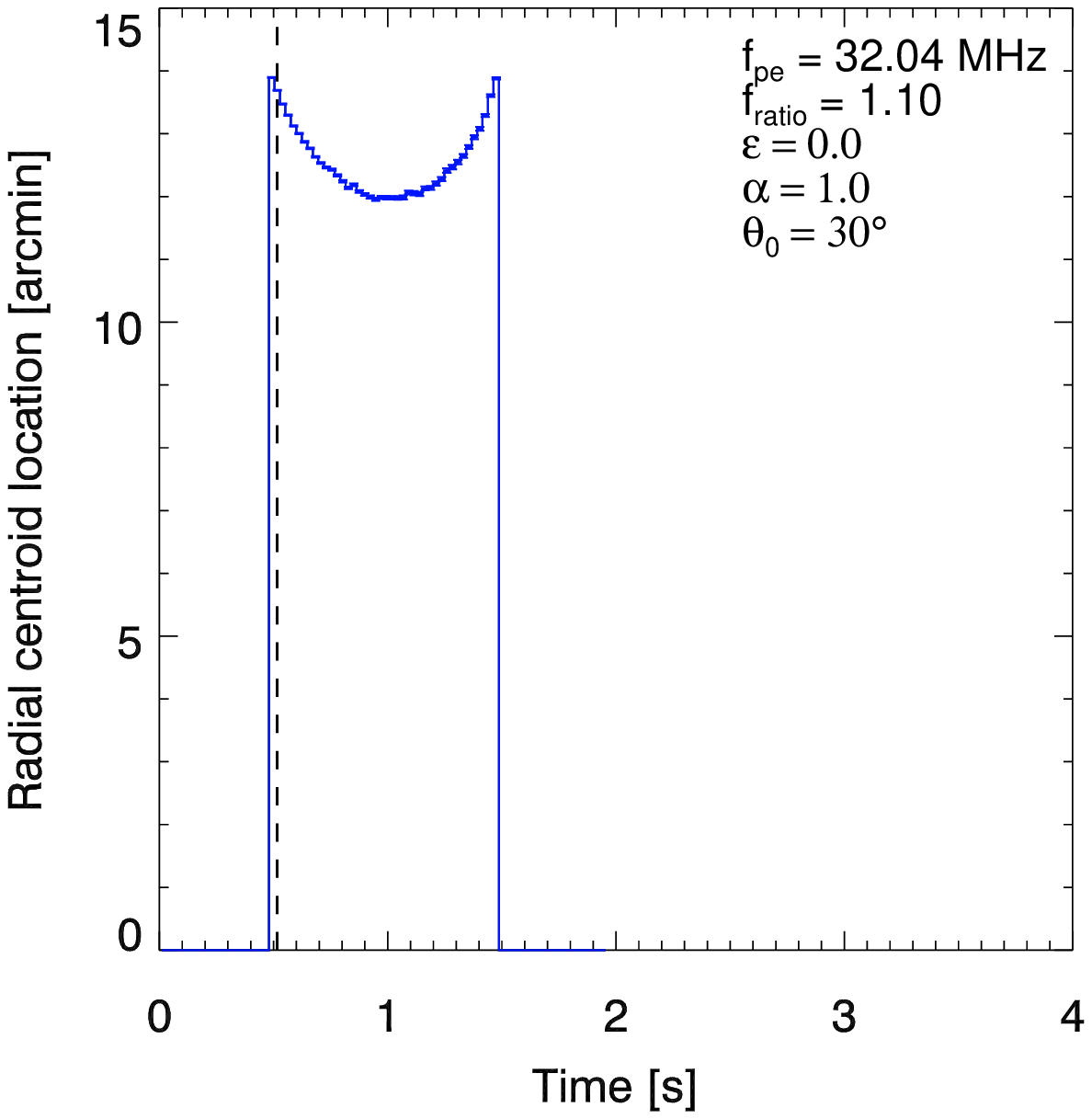}}
\caption{Simulated time profiles without plasma density fluctuations ($\epsilon=0$) for an emission at $f=35.2$ MHz, with $f/f_{\mathrm{pe}}(r_0)=1.10$, and emission sources located at $\theta_0=10^{\circ}$ (left column) and $\theta_0=30^{\circ}$ (right column). Top row: the radio flux (normalized by the maximum value), where the blue and red lines represent the signal with and without the collisional absorption, respectively. Bottom row: the apparent radio source position (distance from the solar disk center), where error bars represent one standard deviation.}
\label{sim_e00}
\end{figure*}

\section{Numerical simulation results}\label{results}
\begin{figure*}
\centerline{\includegraphics[width=0.3\textwidth]{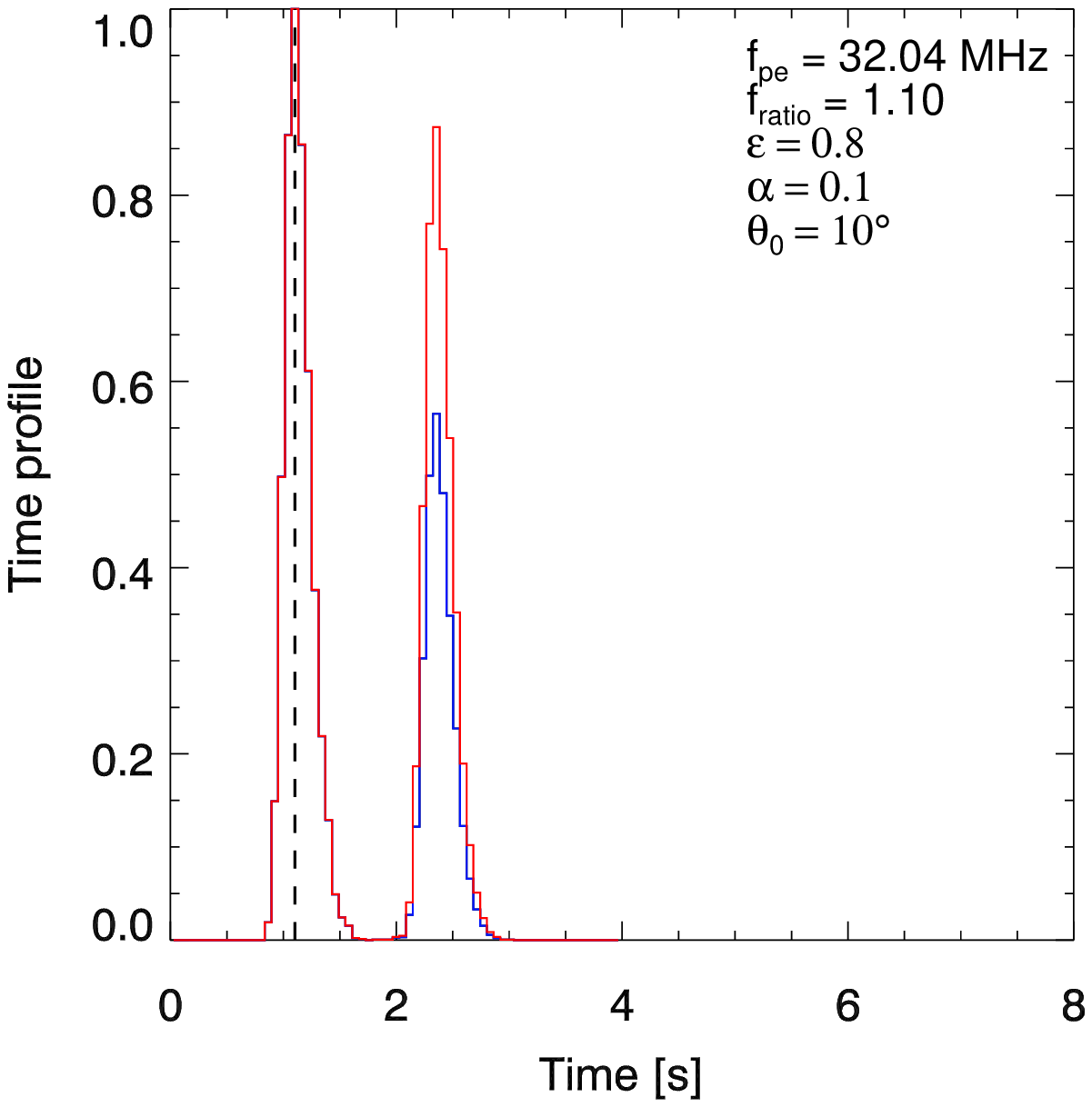}~
\includegraphics[width=0.3\textwidth]{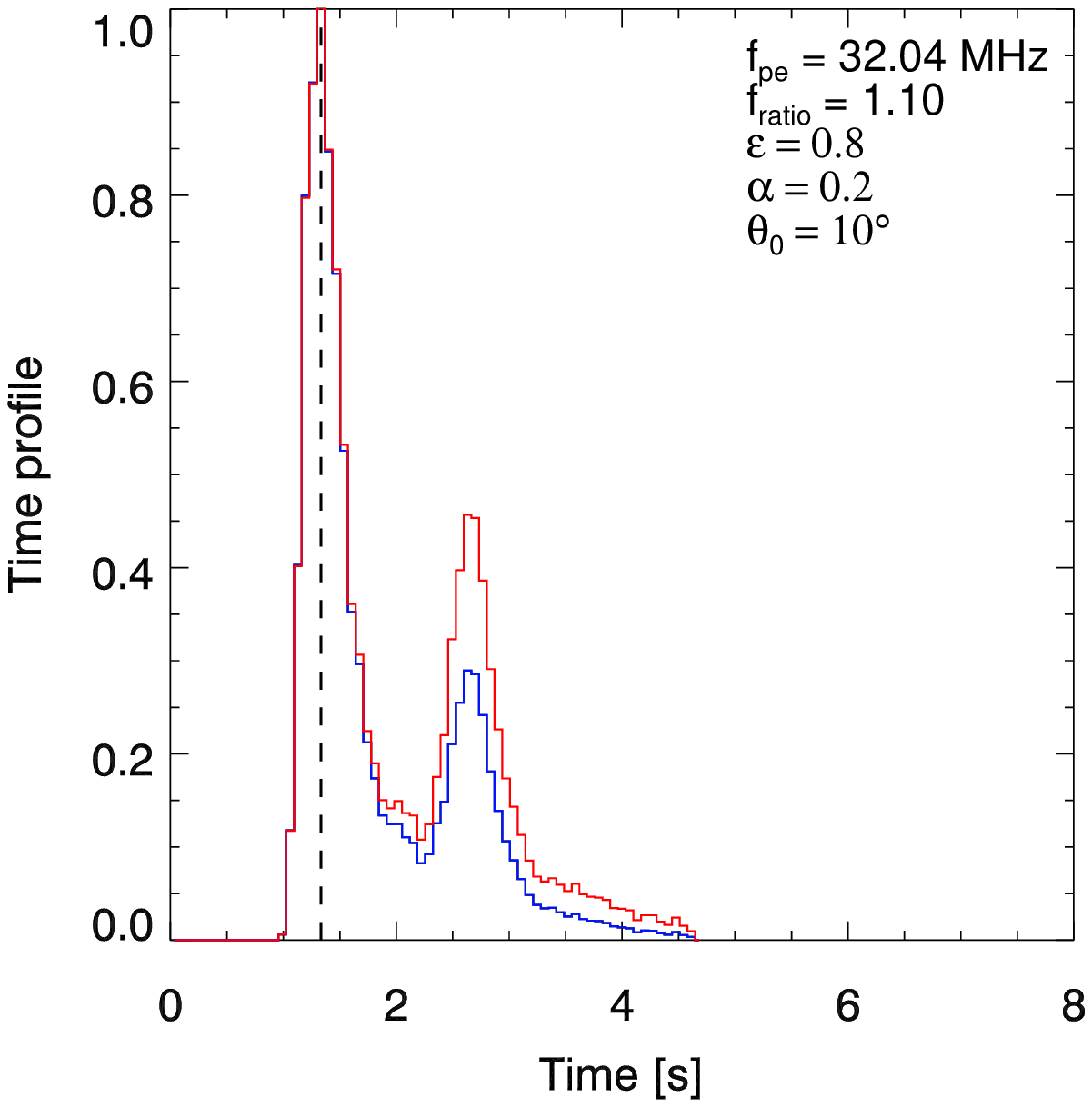}~
\includegraphics[width=0.3\textwidth]{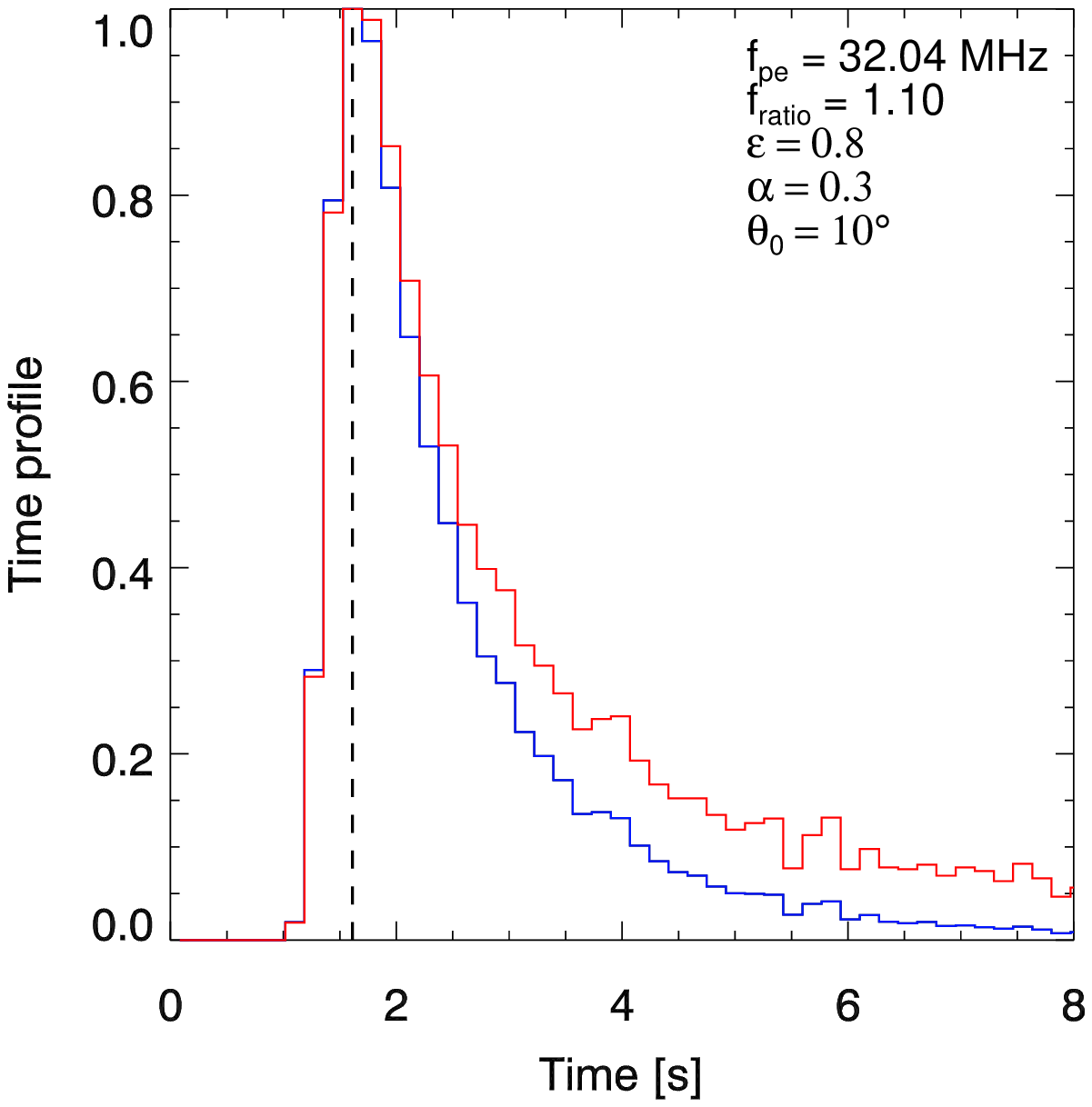}}
\centerline{\includegraphics[width=0.3\textwidth]{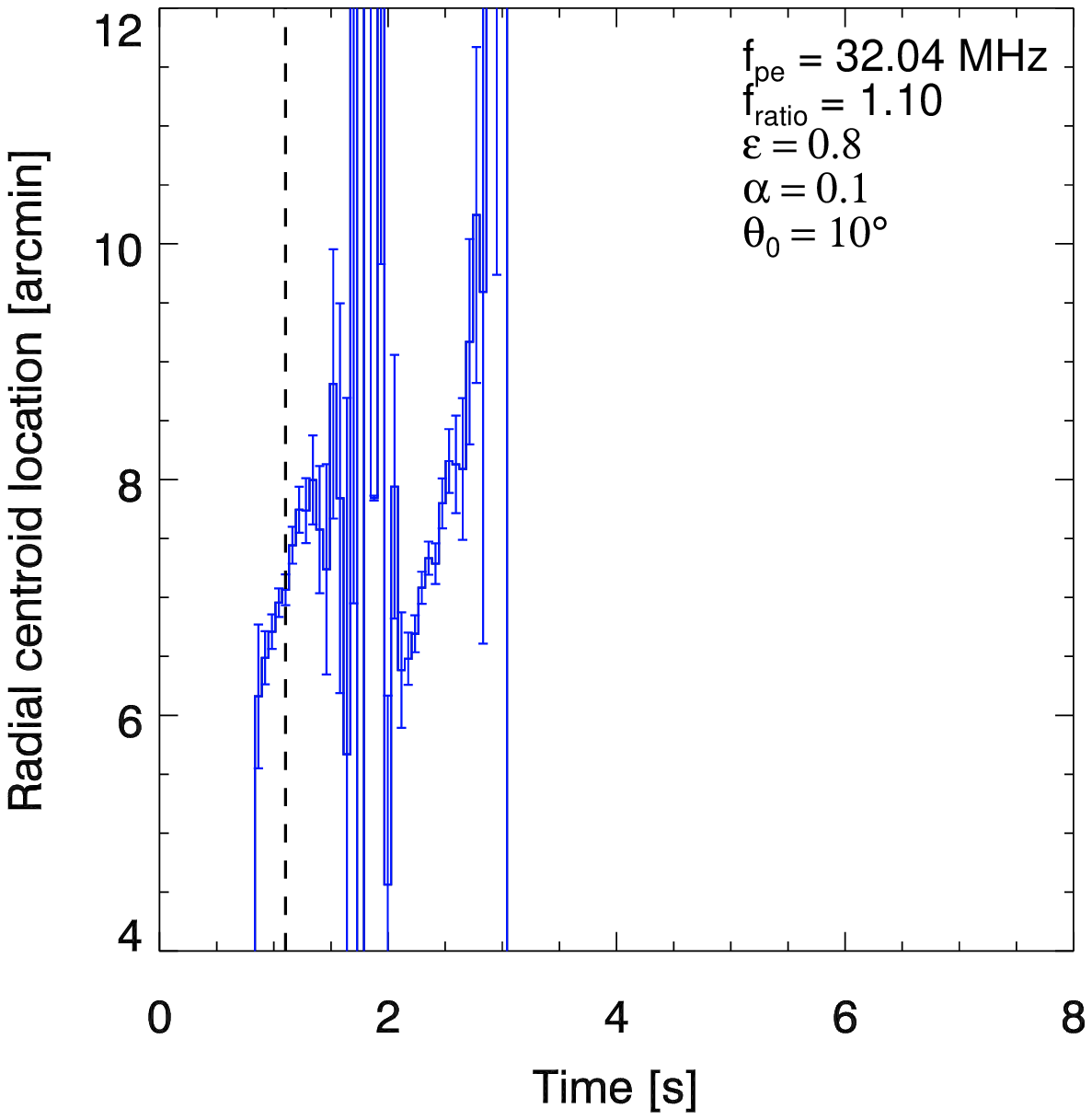}~
\includegraphics[width=0.3\textwidth]{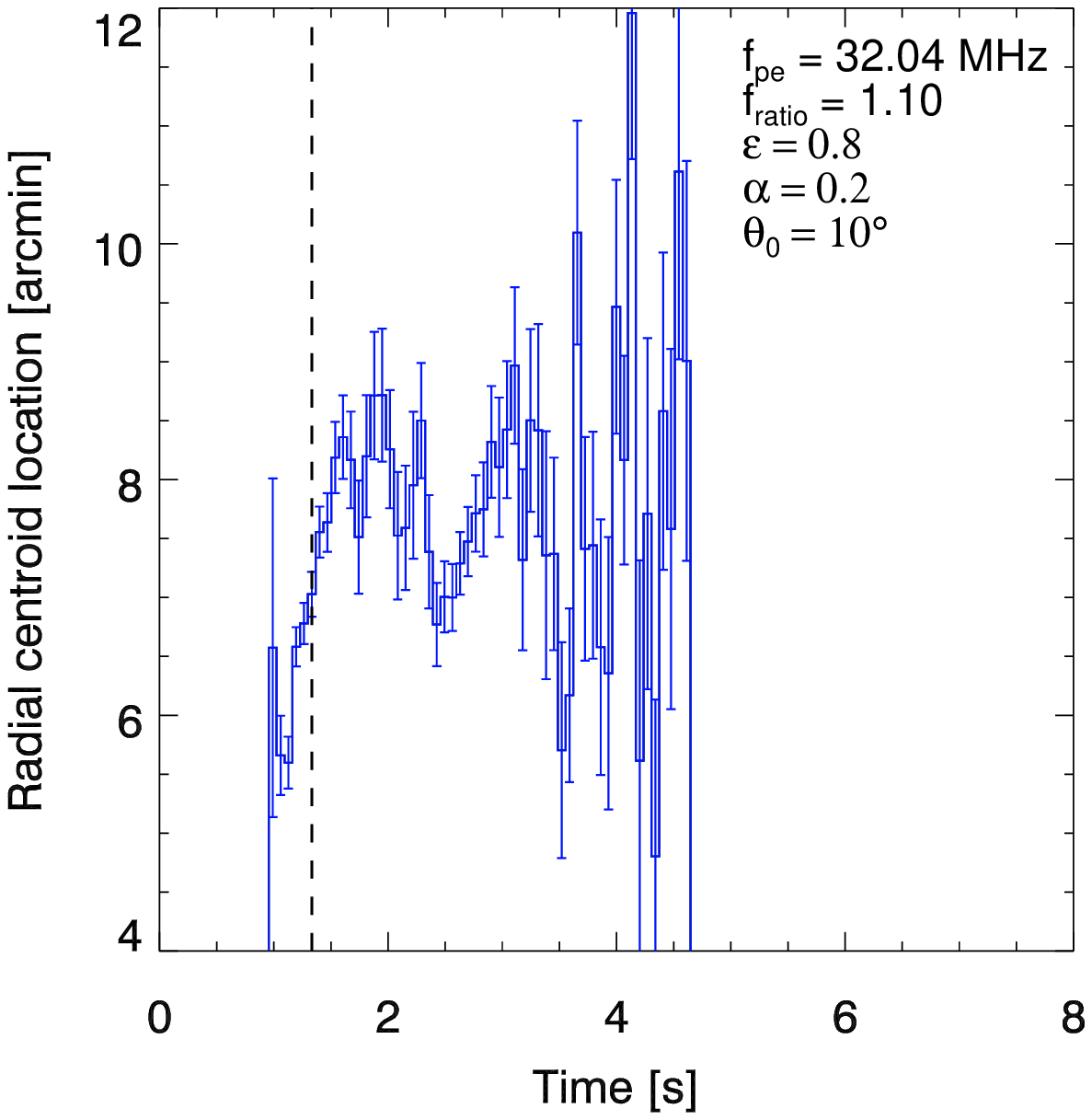}~
\includegraphics[width=0.3\textwidth]{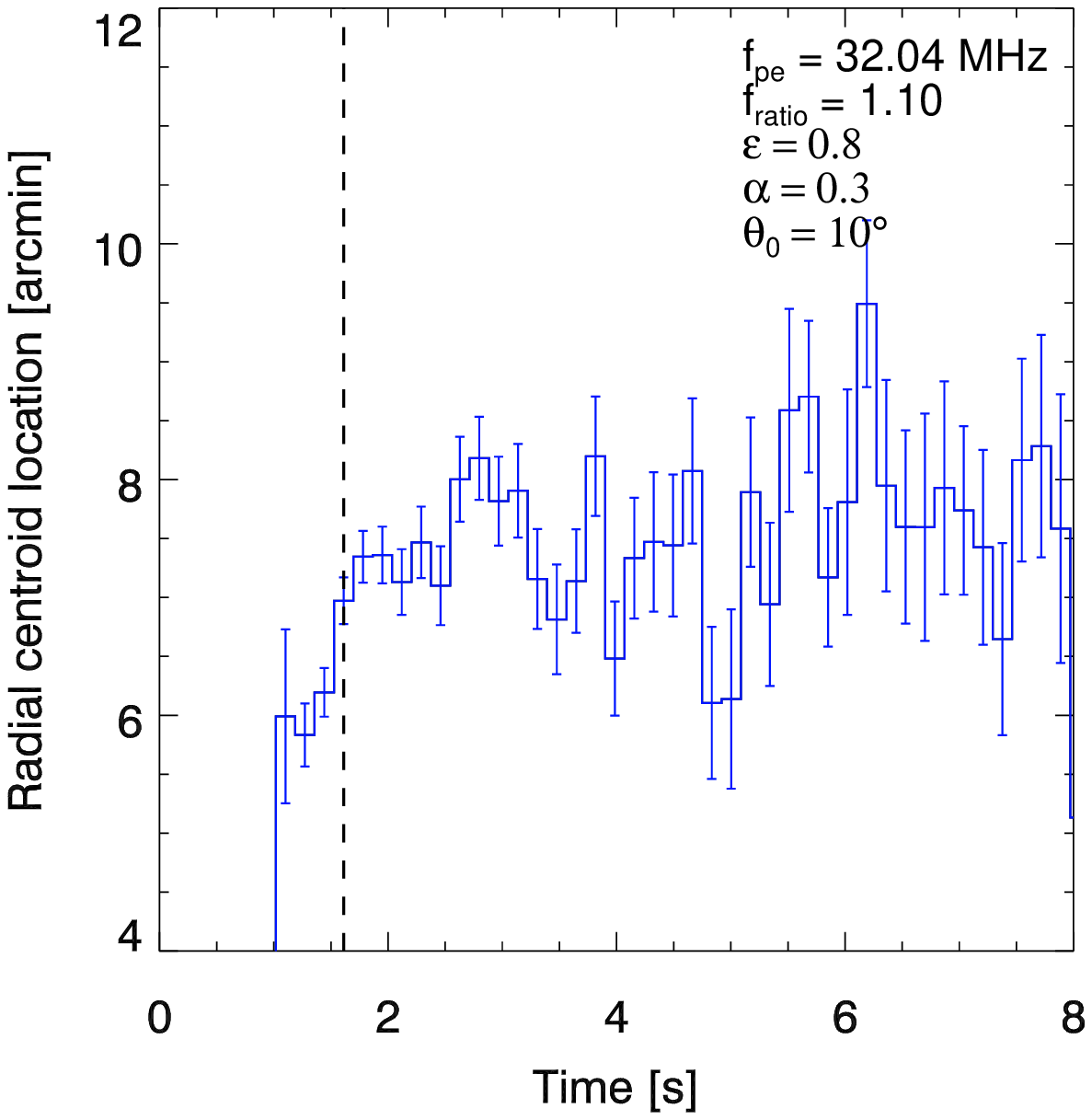}}
\centerline{\includegraphics[width=0.3\textwidth]{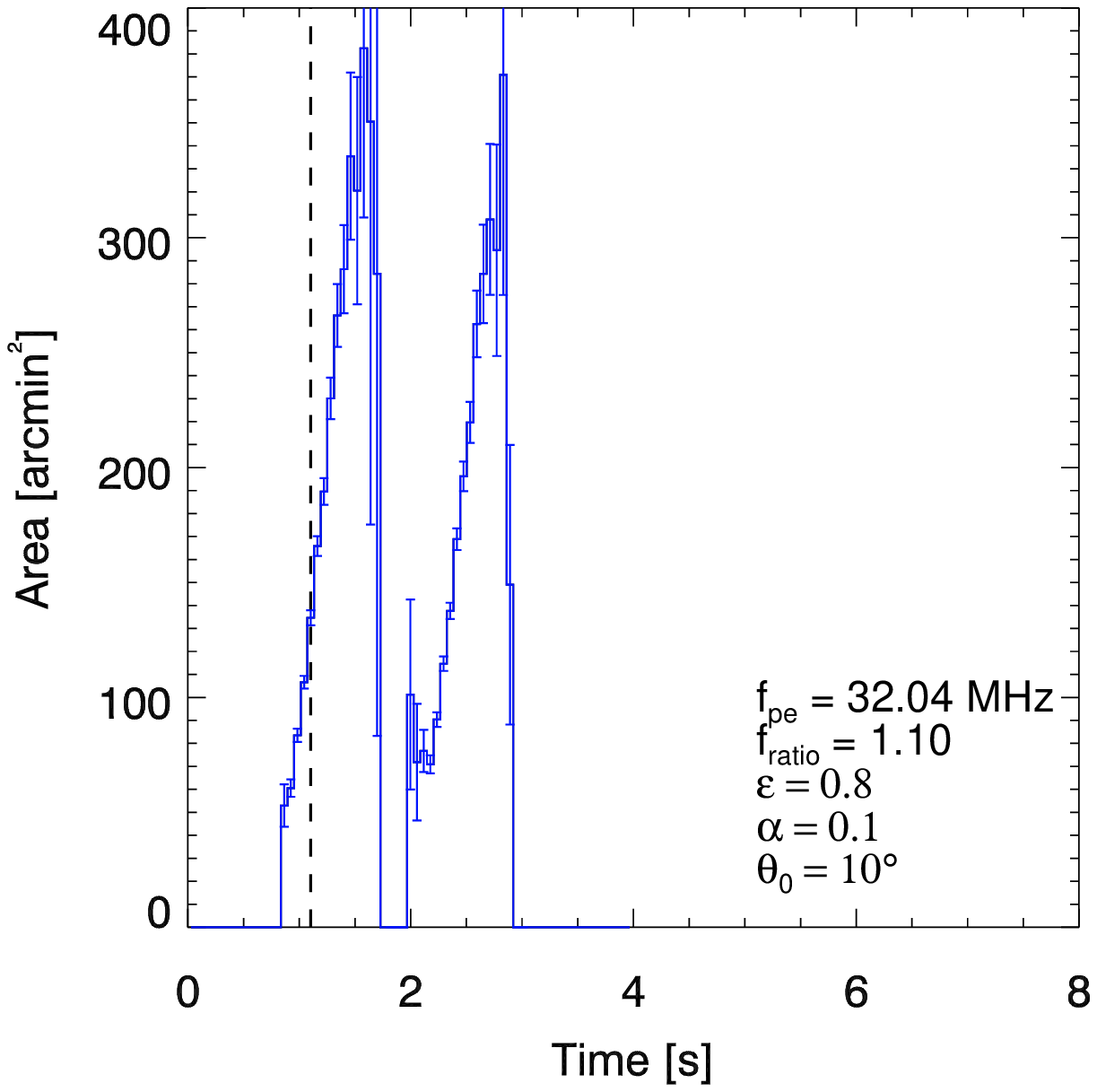}~
\includegraphics[width=0.3\textwidth]{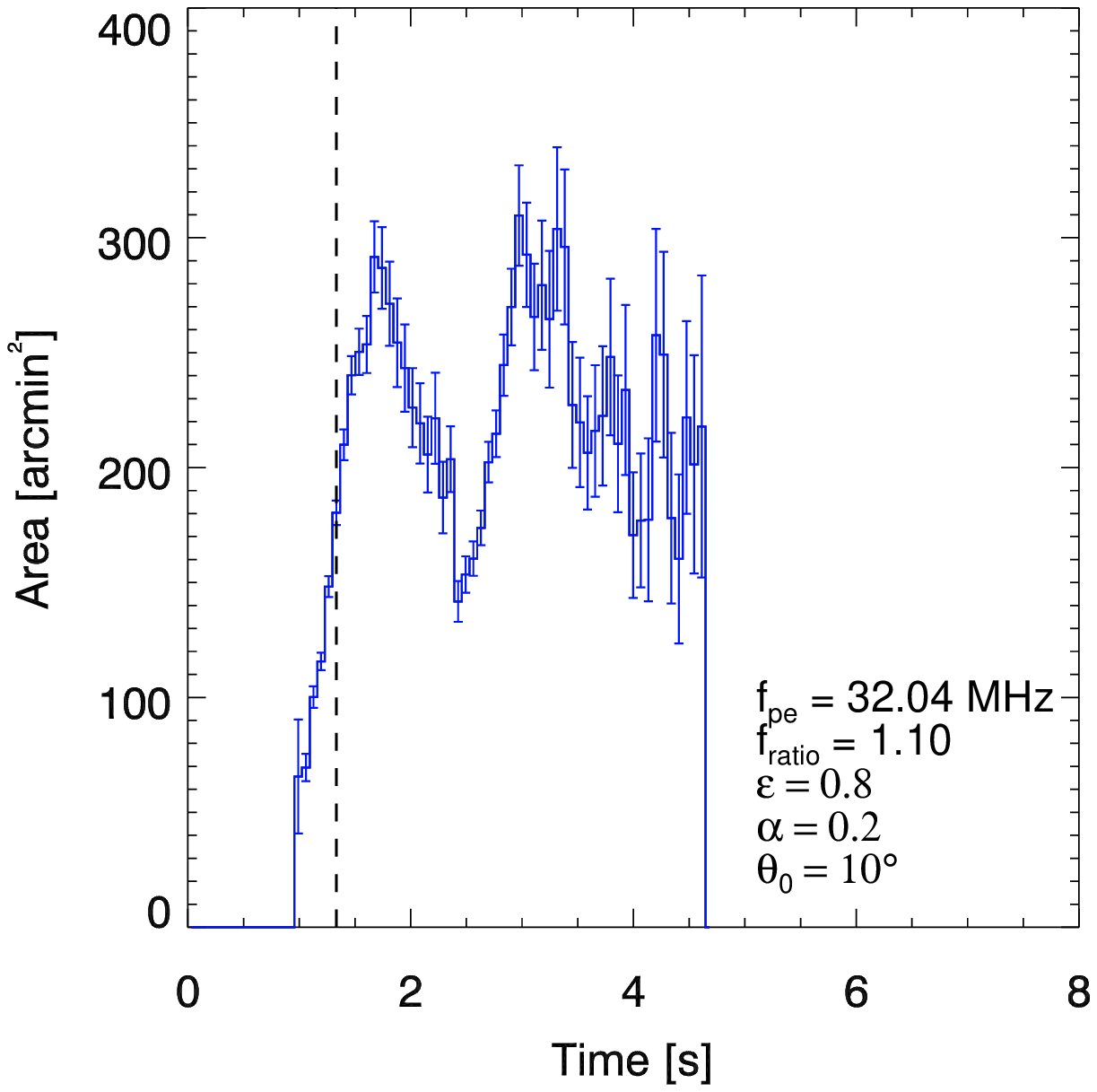}~
\includegraphics[width=0.3\textwidth]{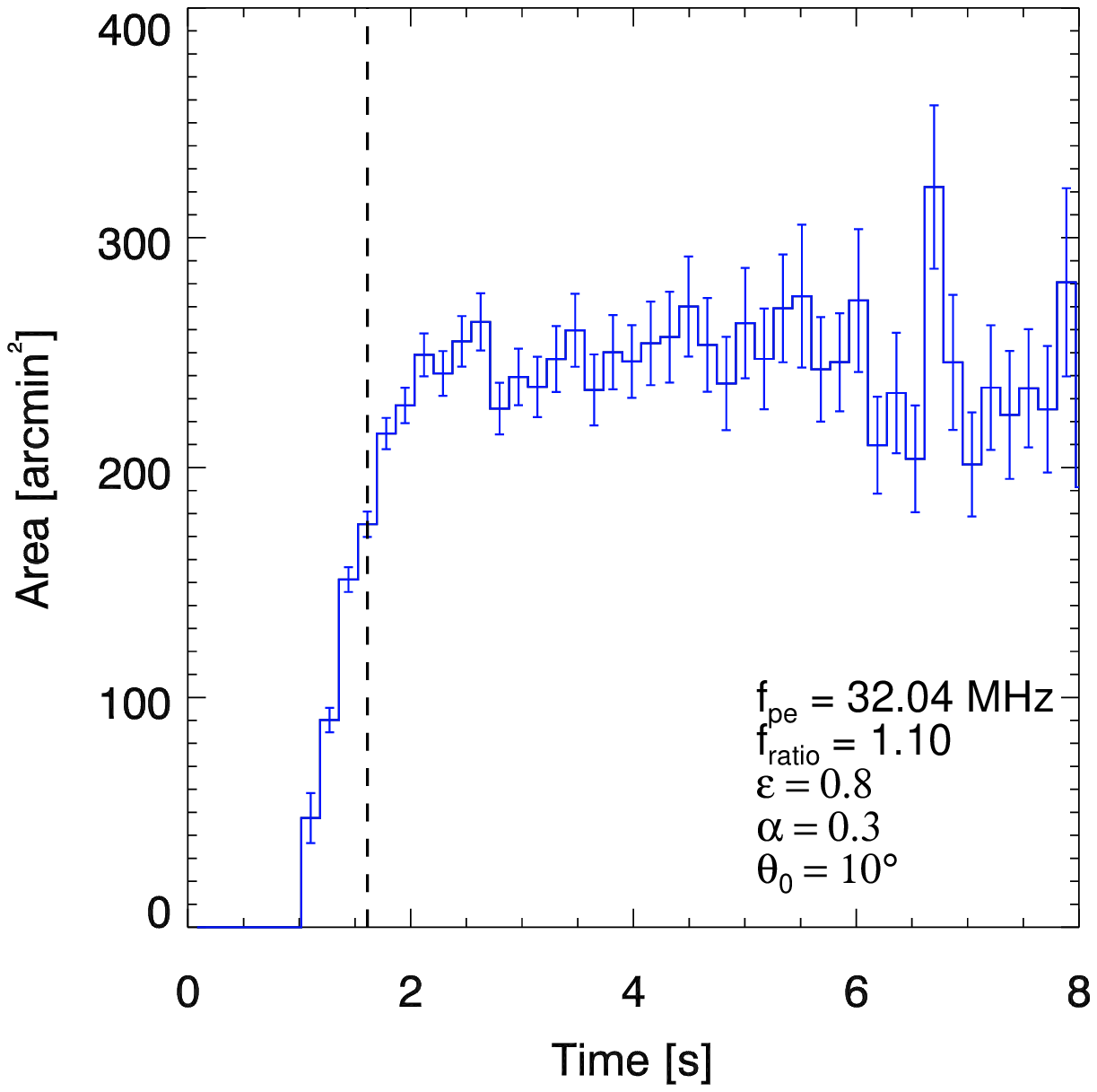}}
\caption{Simulated time profiles for  a level of density fluctuations $\epsilon=0.8$, an emission source located at $\theta_0=10^{\circ}$, an emission frequency of $f=35.2$ MHz, $f/f_{\mathrm{pe}}(r_0)=1.10$, and anisotropy $\alpha=0.1$ (left column), $0.2$ (middle column), and $0.3$ (right column). Top row: the radio flux (normalized by the maximum value) where the blue and red lines represent the signal with and without the collisional absorption, respectively. Middle row: the apparent radio source position (distance from the solar disk center). Bottom row: the apparent radio source area (at half-maximum level). Error bars represent one standard deviation.}
\label{sim_alpha}
\end{figure*}

Figures \ref{sim_e00}--\ref{sim_theta} demonstrate the simulated time profiles of the emission intensity and the apparent radio source position and size for several representative combinations of parameters. We note that due to the finite number of photons, the calculated parameters will have a statistical error. In particular, the source position and source size have lower uncertainties when the number of photons is larger (i.e. near the peak of the light curves), while the calculated values away from the peak have larger errors. Another important factor to consider is that, in contrast to real observations, our simulations do not include background radio sources, which complicates real observations (see Figure \ref{obs_prof}).

\subsection{  Reflection/refraction without scattering}\label{noscattering}
We first examine the case when the plasma density fluctuations are absent ($\epsilon=0$) and the radio-wave propagation is determined entirely by the refraction and reflection processes. Earlier models \citep{roberts_1958} suggested that the second elements of the bursts are echoes of the first ones reflected from lower levels of the solar corona. Figure \ref{sim_e00} shows the time profiles of the emission intensity and apparent source position for the emission frequency of $f=35.2$ MHz, the initial emission to local plasma frequency ratio of $f/f_{\mathrm{pe}}(r_0)=1.10$, and the emission source located at the heliocentric longitudes of $\theta_0=10^{\circ}$ and $30^{\circ}$. For the source located at $\theta_0=10^{\circ}$ (i.e., rather close to the solar disk center, as the observations indicate), the light curve demonstrates a sharp decay and then a very weak secondary peak (reflected component) delayed by $\sim 0.7$ s with respect to the first one. For the source located at $\theta_0=30^{\circ}$, the decay is slower and the secondary (reflected) component is lost in the tail of the first one. The temporal evolution of the apparent emission source position differs from the observed one (cf. Figure \ref{obs_prof}c) even more dramatically: the source firstly shifts towards the disk center and then bounces back. This reflection is reminiscent of X-ray scattering in the lower atmosphere \citep{jeffrey_2011}, but with account for refraction, so that the radio waves, unlike X-rays, do not propagate along straight lines and can be reflected at various (time-dependent) locations. The apparent source locations of the first and second components nearly coincide for the viewing angle of $\theta_0=10^{\circ}$. For the fundamental emission with $f/f_{\mathrm{pe}}(r_0)=1.10$ and the frequency of about 30 MHz, the projected distance between the emission source and the nearest reflection point is about $0.75\sin\theta_0$ arcmin. The apparent source areas (not shown in the figure) do not exceed 5 $\textrm{arcmin}^2$, which are much smaller than the observed ones. Thus we conclude that the pure reflection and refraction model (either fundamental or harmonic) is not sufficient to reproduce the observed drift-pair burst properties. On the other hand, the model involving the fundamental plasma emission is somewhat better for explaining nearly coinciding source locations of the direct and reflected components.

\subsection{Echo in anisotropic scattering media}
We now consider the model where the plasma density fluctuations are present ($\epsilon=0.8$). In Figure \ref{sim_alpha}, we demonstrate the effect of the anisotropy level $\alpha$ which we vary from $0.1$ to $0.3$. The emission frequency is taken to be $f=35.2$ MHz, the initial emission to local plasma frequency ratio is $f/f_{\mathrm{pe}}(r_0)=1.10$, and the emission source is located at the heliocentric longitude $\theta_0=10^{\circ}$. For a high anisotropy level of $\alpha=0.1$ (left column in Figure \ref{sim_alpha}), the light curve has an evident double-peak structure; the second (reflected) component is delayed by $\sim 1.25$~s with respect to the first (direct) one. Both peaks are rather short in duration ($\sim 0.5$ s). The second component has a slightly lower amplitude (note the effect of collisional absorption); nevertheless, the component amplitudes are comparable and the overall shape of the simulated light curve agrees well with the observations (cf. Figure \ref{obs_prof}b). Notably, the apparent sources of both components coincide spatially: they are located (if we consider the intensity peaks) at the same distance of $\sim 7.0$ arcmin from the solar disk center. The sources move outwards with the rate of about $4.0$ arcmin $\textrm{s}^{-1}$. The apparent source areas of both components are also nearly identical  ($\sim 140$ $\textrm{arcmin}^2$ at the intensity peaks), and the sources expand at a rate of about 520 $\textrm{arcmin}^2$ $\textrm{s}^{-1}$. The simulated source position and size (with correction for the LOFAR beam size) agree well with the observations. On the other hand, the simulated source motion and (especially) the expansion rates are considerably higher than the observed ones. This discrepancy may be attributed to the fact that, in contrast to simulations, in real observations we measure the centroid location and effective size of a combined source including the contributions of a variable bursty signal and a background continuum (i.e., we obtain a weighted average of the locations and sizes of the corresponding sources), which reduces the resulting variation rates of the source parameters.

For a lower anisotropy level $\alpha=0.2$ (see middle column in Figure \ref{sim_alpha}), the light curve still has a double-peak structure. However, the peaks become broader (with a FWHM duration of $\sim 0.7$ s), and the delay between the components slightly increases (up to $\sim 1.30$ s). The most important difference from the previous case with $\alpha=0.1$ is that the relative amplitude of the second (reflected) component decreases considerably. The change of the anisotropy level has almost no effect on the apparent radio source position: the sources of both components (at the intensity peaks) are located at the same distance of $\sim 7.0$ arcmin from the solar disk center. The apparent source size increases slightly (to $\sim 150$ $\textrm{arcmin}^2$ at the intensity peaks), but the source expansion rate decreases considerably (down to $\sim 370$ $\textrm{arcmin}^2$ $\textrm{s}^{-1}$).

For even lower anisotropy levels (e.g., $\alpha=0.3$; see right column in Figure \ref{sim_alpha}), light curves of both the direct and reflected components become broader, so that the trailing component is almost invisible (i.e., the reflected component is lost in the tail of the first component).  The source size at the peak of the burst is about 180 $\textrm{arcmin}^2$. Therefore the formation of drift-pair bursts requires a sufficiently strong anisotropy of the plasma turbulence ($\alpha\lesssim 0.1-0.2$).

The apparent sources of drift-pair bursts are more compact than those of type III bursts at the same frequency: $\sim 150$ $\textrm{arcmin}^2$ vs. $\sim 350$ $\textrm{arcmin}^2$, respectively, at 30 MHz, after correction for the instrument's beam size \citep[see][]{suzuki_1979, dulk_1980, kuznetsov_2019}. According to the presented simulations, the apparent source size increases when the plasma turbulence becomes more isotropic. This increase, however, is not enough to explain the observed difference in the source sizes of different burst types. Therefore, the larger apparent source sizes of type III bursts are likely caused by a higher turbulence level (or, more accurately, a higher scattering rate proportional to $\overline{q\epsilon^2}$) in and around the emission sources: e.g., according to simulations of \citet{kontar_2019}, the apparent source area is growing proportional to $\epsilon$ for the same fluctuation scales $\overline{q}$.

\begin{figure*}
\centerline{\includegraphics[width=0.3\linewidth]{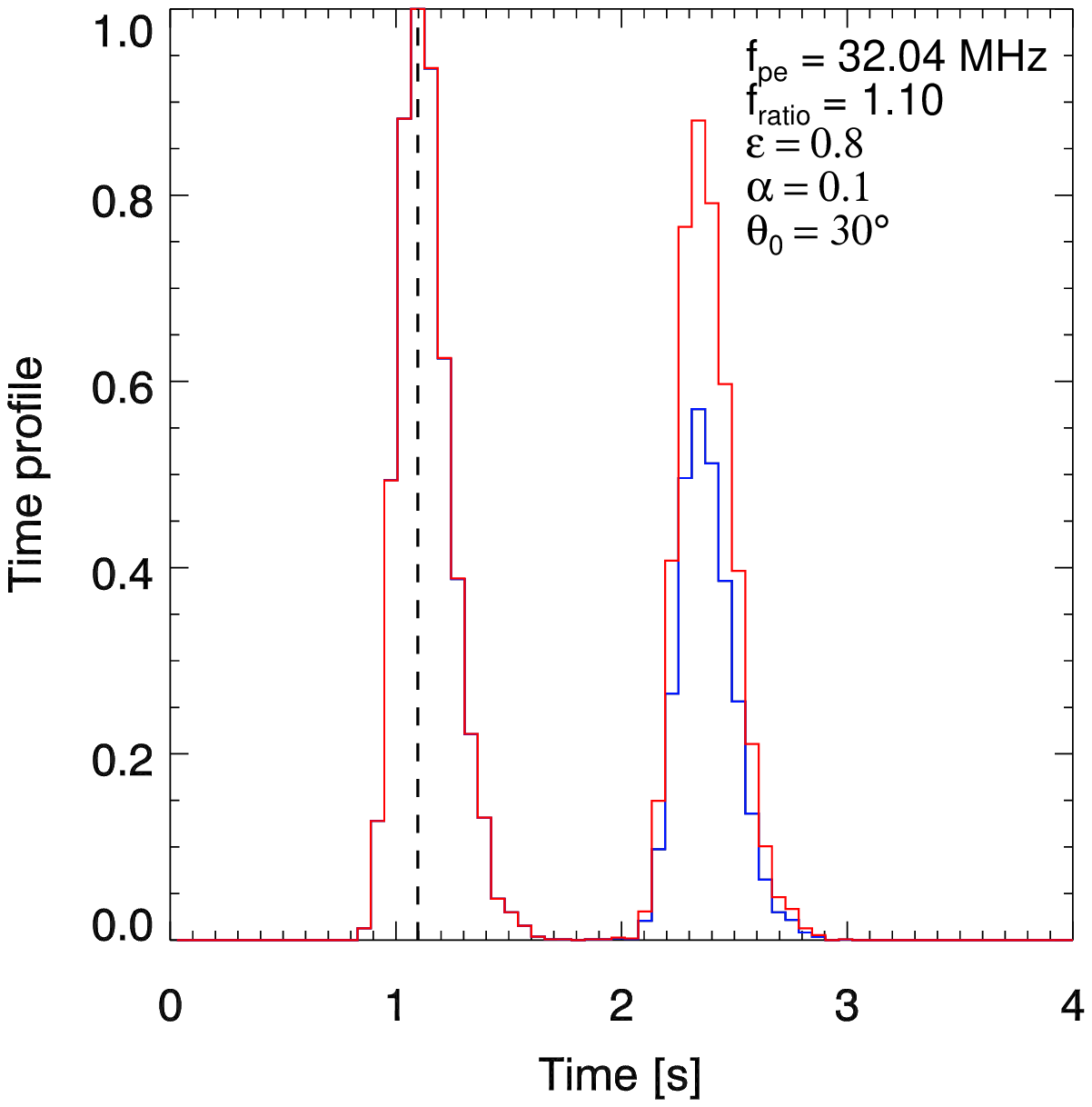}~
\includegraphics[width=0.3\linewidth]{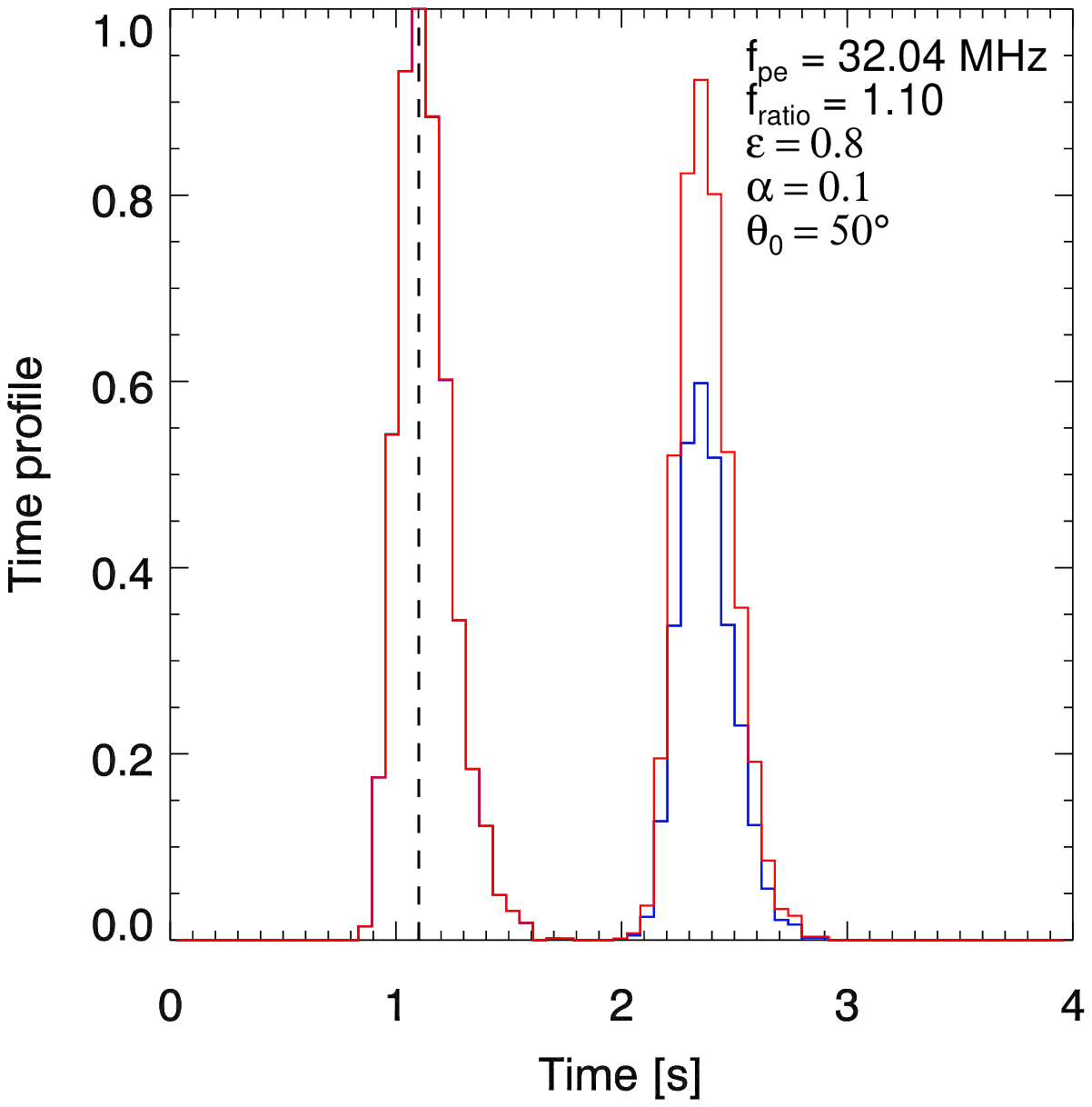}}
\centerline{\includegraphics[width=0.3\linewidth]{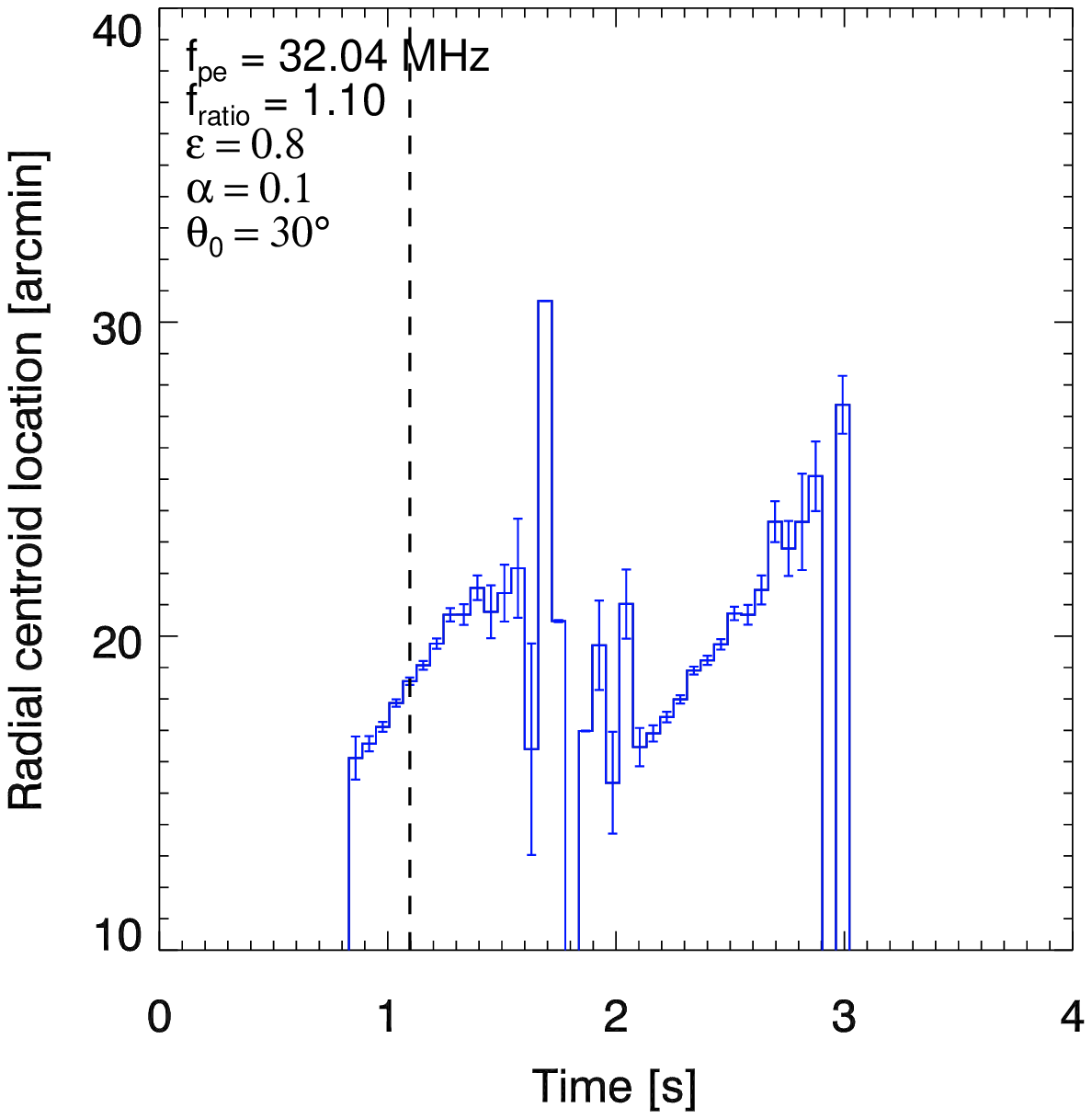}~
\includegraphics[width=0.3\linewidth]{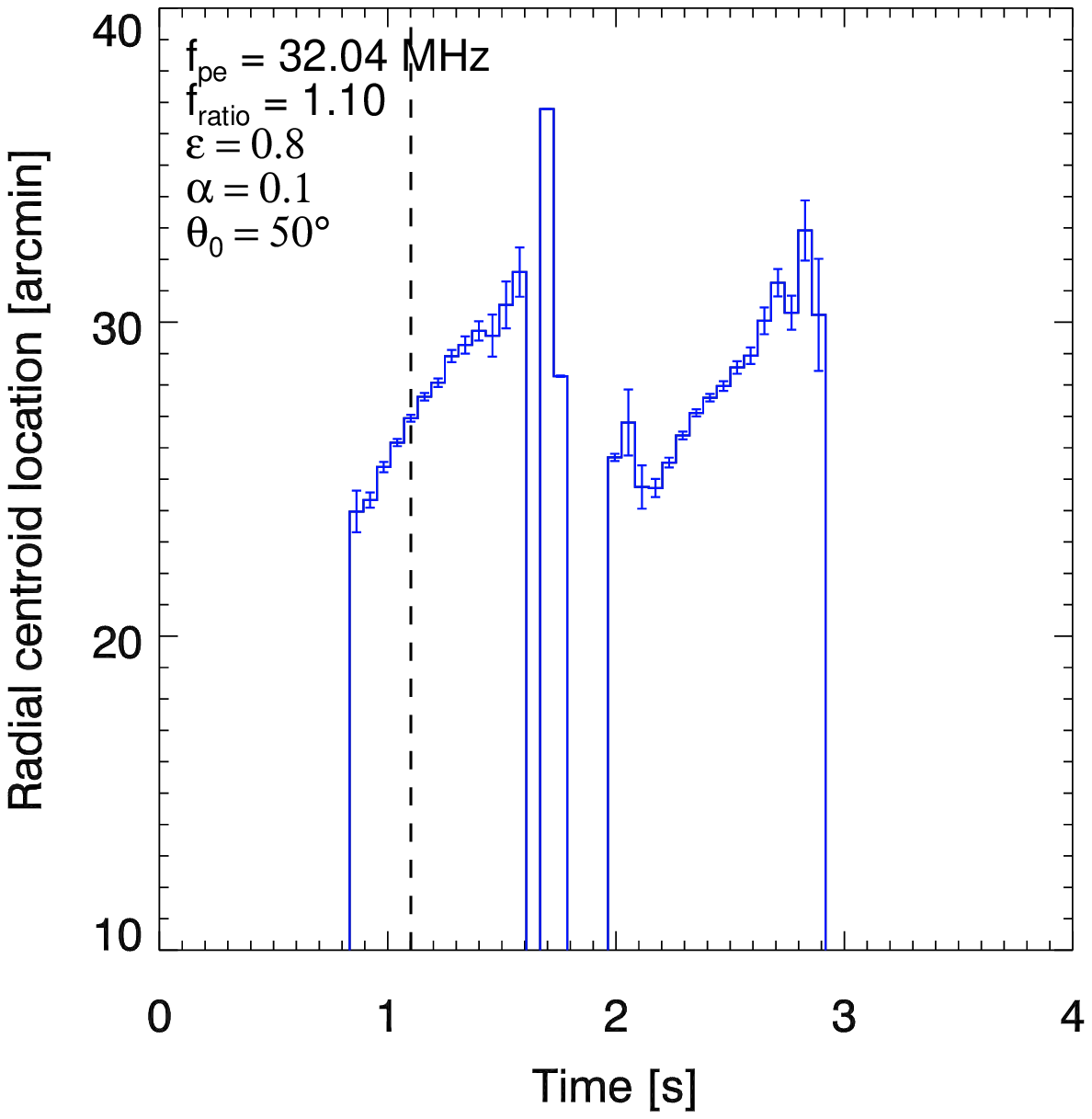}}
\centerline{\includegraphics[width=0.3\linewidth]{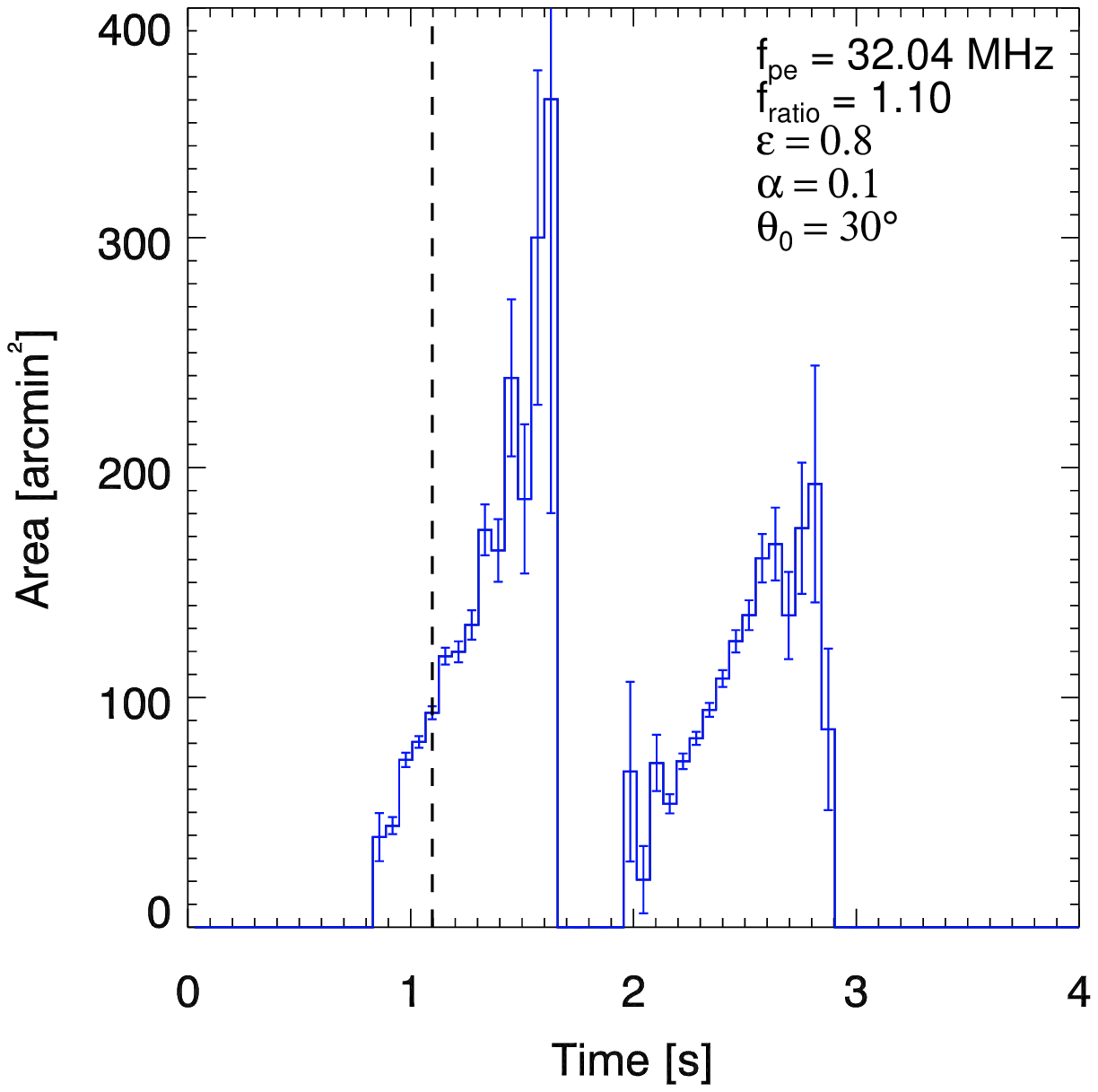}~
\includegraphics[width=0.3\linewidth]{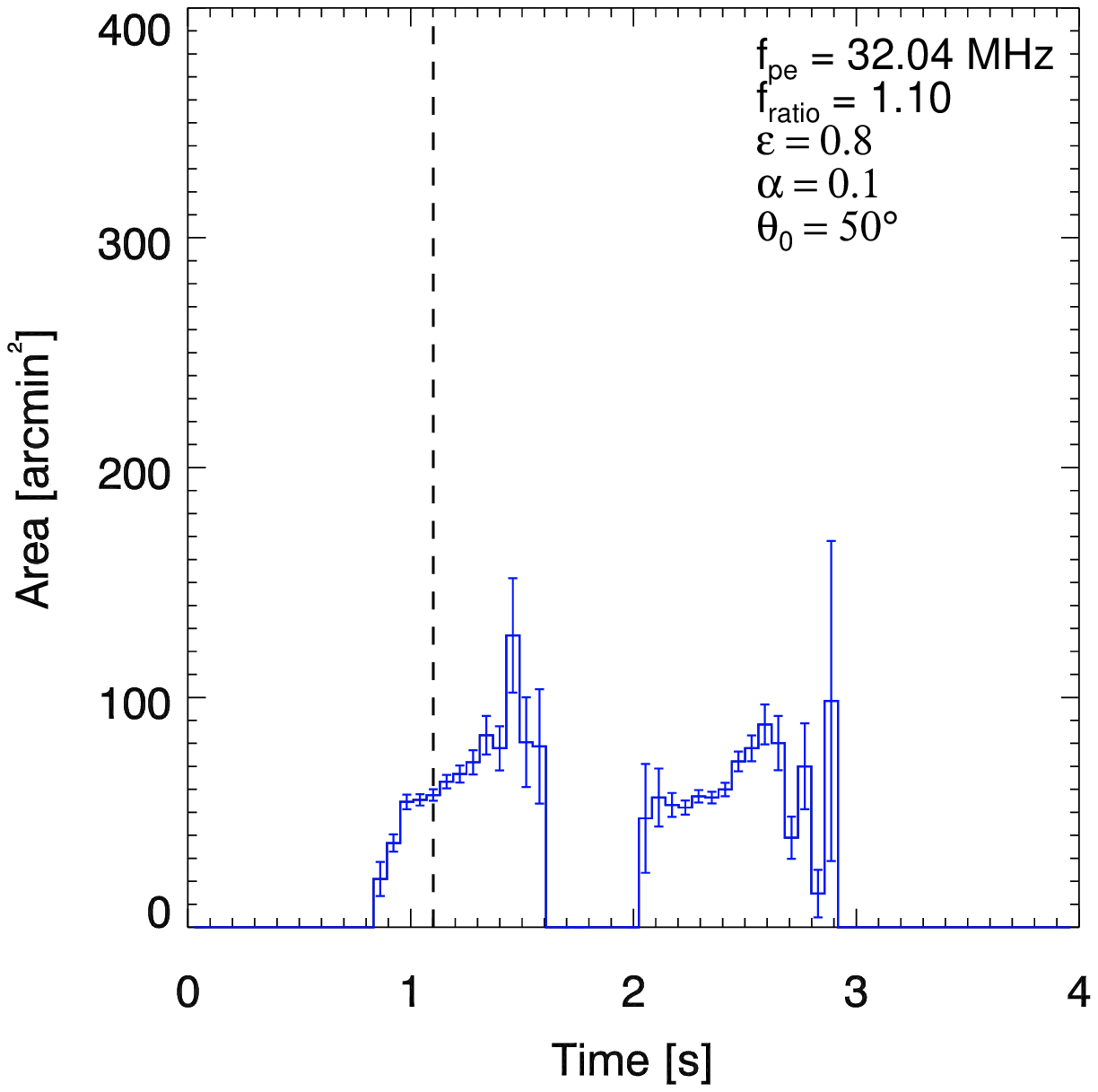}}
\caption{Same as in Figure \protect\ref{sim_alpha}, for  $\epsilon=0.8$, $\alpha=0.1$, emission at $f=35.2$ MHz, $f/f_{\mathrm{pe}}(r_0)=1.10$, and emission sources located at $\theta_0=30^{\circ}$ (left column) and $\theta_0=50^{\circ}$ (right column).}
\label{sim_theta}
\end{figure*}

\begin{figure*}
\centerline{\includegraphics[width=0.3\textwidth]{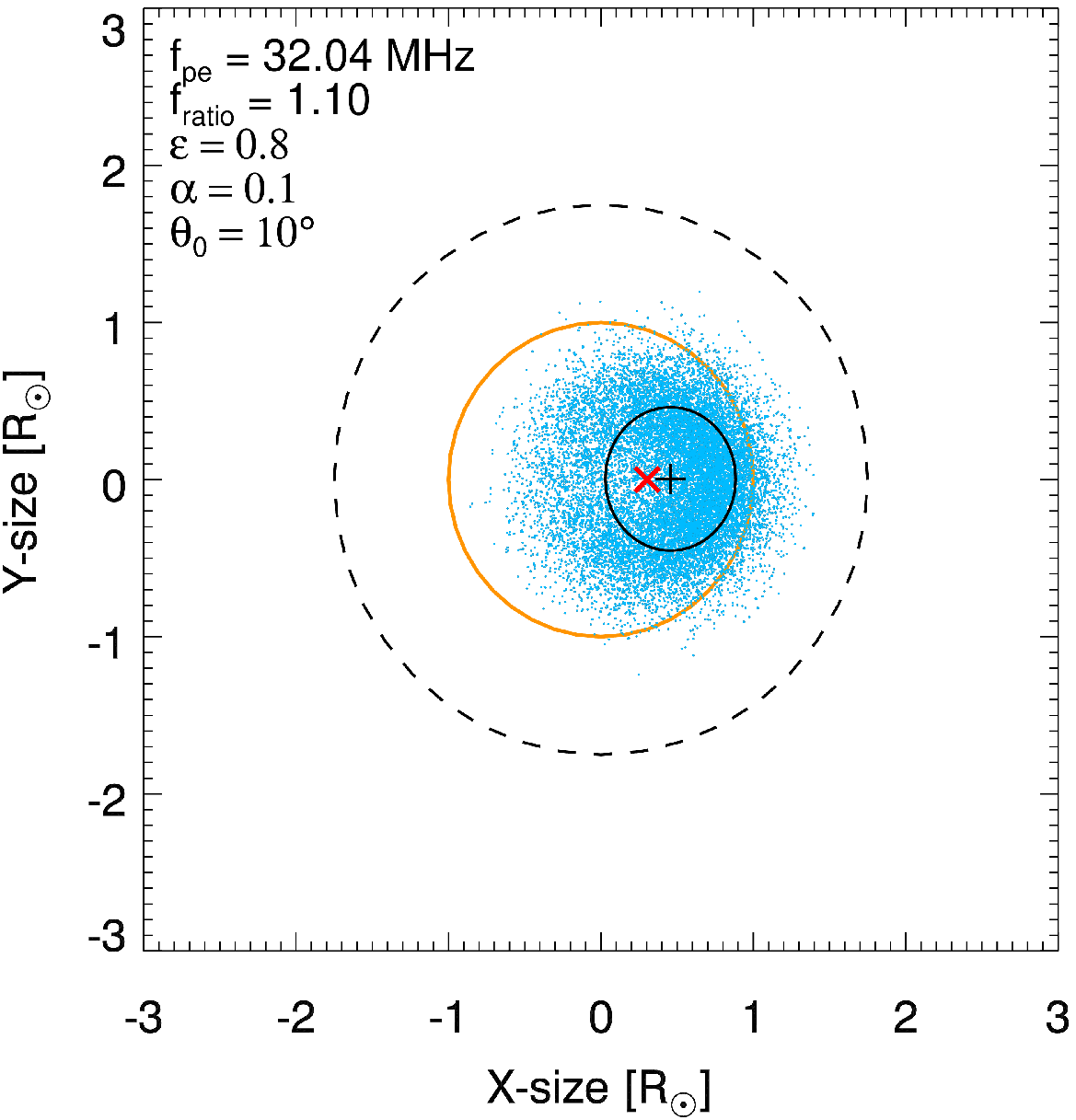}~
\includegraphics[width=0.3\textwidth]{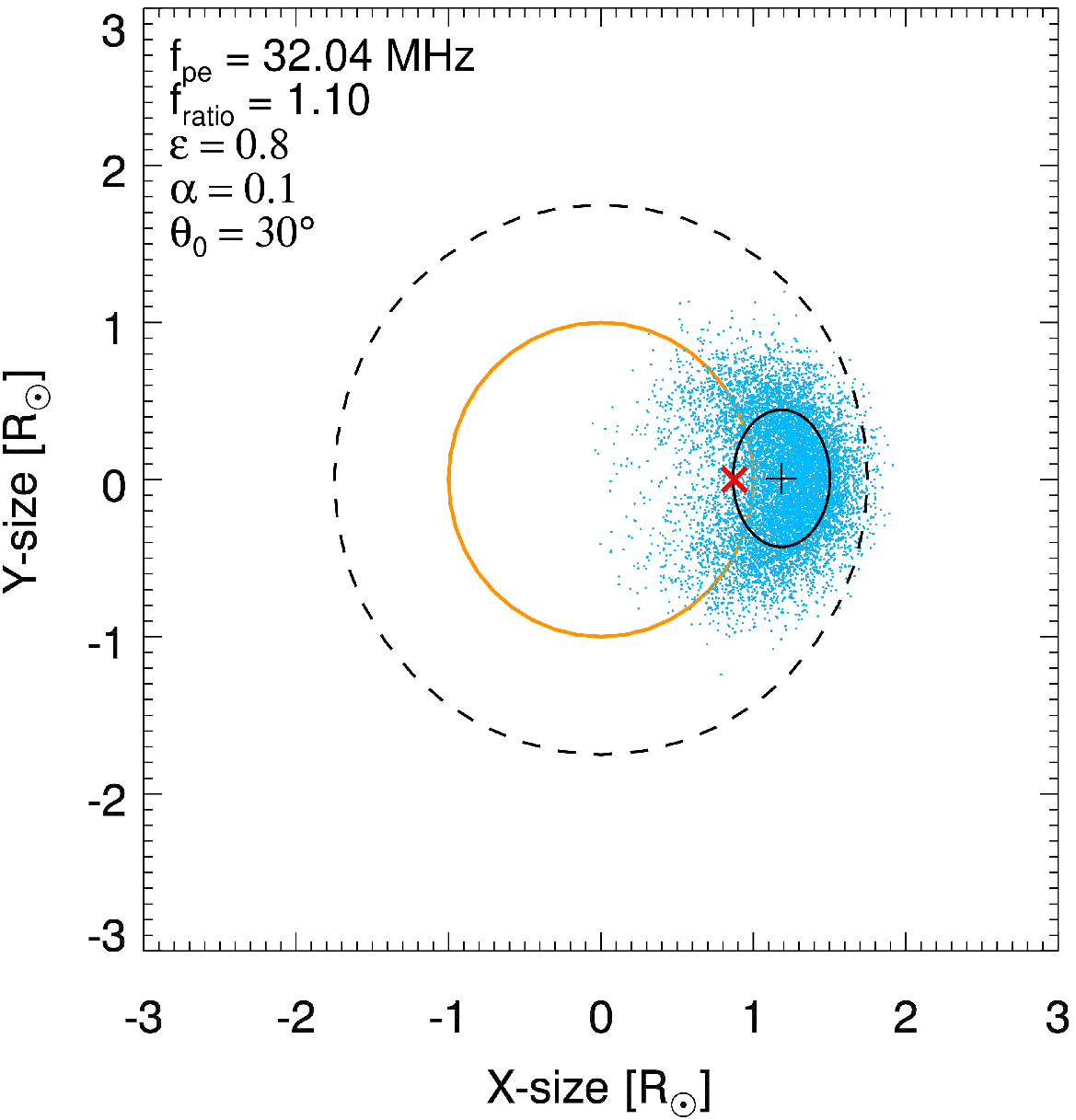}~
\includegraphics[width=0.3\textwidth]{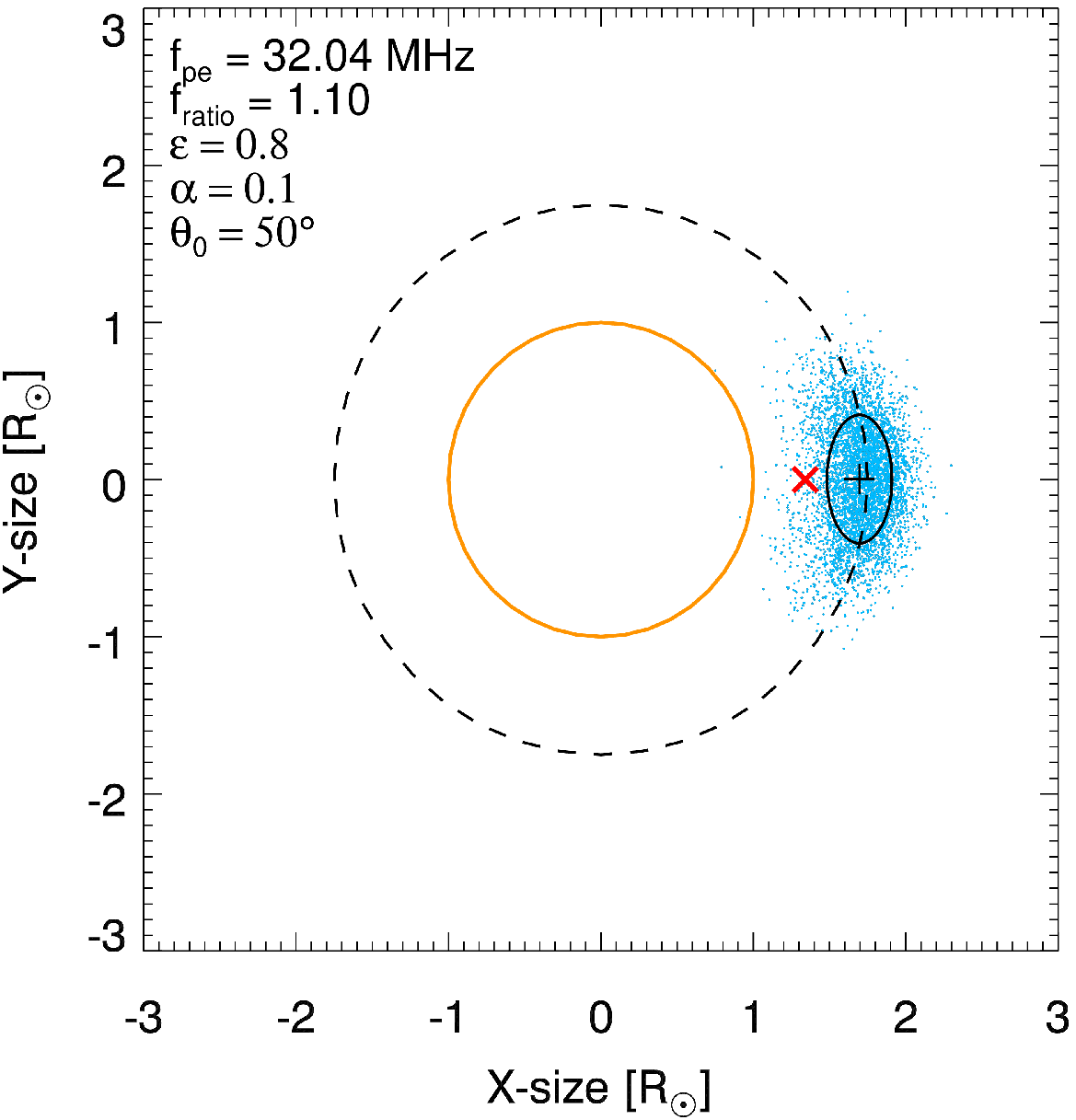}}
\caption{Simulated radio images for emission at $f=35.2$ MHz, $f/f_{\mathrm{pe}}(r_0)=1.10$, a density fluctuation level $\epsilon=0.8$, a level of anisotropy $\alpha=0.1$, and an emission source located at longitudes $\theta_0 = 10^{\circ}$, $30^{\circ}$, and $50^{\circ}$. Each dot represents one photon. The projected position of the true radio source and the apparent source centroid are shown by the red ($\times$) and black ($+$) crosses, respectively. The black ellipse shows the source area at the half-maximum level. The orange circle denotes the solar limb and the dashed circle denotes the level where the true source is located.}
\label{sim_img}
\end{figure*}

\subsection{Center-to-limb variation}
Figures \ref{sim_img}, \ref{sim_theta} show the simulated time profiles for different initial source locations (for $\epsilon=0.8$, $\alpha=0.1$, emission at $f=35.2$ MHz, and $f/f_{\mathrm{pe}}(r_0)=1.10$). In combination with the first column of Figure \ref{sim_alpha}, Figure \ref{sim_theta} covers the range of source heliocentric longitudes $\theta_0$ from $10^{\circ}$ to $50^{\circ}$. For all source locations, the radio light curves demonstrate a very similar double-peak structure with the same delay ($\sim 1.25$ s) between the components. As expected, the apparent source position is strongly dependent on its true position; however, in all cases the apparent sources of the first and second burst components coincide spatially, with a good accuracy. The source coincidence is caused both by the fact that the emission is produced at the fundamental plasma frequency (and hence, as mentioned in Section \ref{noscattering}, the projected distance between the emission source and the nearest reflection point is less than 1 arcmin for the considered parameters), and by the effect of scattering which makes the emission directivity pattern narrower and thus restricts the possible range of trajectories for the observable radio waves. The apparent source motion speed reaches its maximum ($\sim 10$ arcmin $\textrm{s}^{-1}$) at longitudes $\theta_0 \simeq 30^{\circ}-50^{\circ}$. Both the apparent source size and expansion rate gradually decrease (from $\sim 180$ to $\sim 70$ $\textrm{arcmin}^2$ and from $\sim 520$ to $\sim 50$ $\textrm{arcmin}^2$ $\textrm{s}^{-1}$, respectively) when the source shifts away from the solar disk center.

The above trends are also visible in the radio brightness maps shown in Figure \ref{sim_img} (the figure shows time-integrated maps, which include both the direct and reflected burst components). Due to refraction and scattering, the apparent radio source is located farther from the solar disk center than the true source. This shift increases with distance from the disk center, since the effect of scattering becomes more pronounced. A notable feature is that both the apparent source size and the emission intensity (i.e., number of photons reaching the Earth) decrease with increasing distance from the disk center. Therefore, the locations near the solar disk center are preferable because they would provide a higher radio flux (including higher signal-to-noise and signal-to-background ratios) and therefore a higher probability to detect the bursts. On the other hand, anisotropic scattering and refraction are able to produce the characteristic double-peak light curve (with spatially coinciding sources of both components) for all source locations. Comparing the simulation results (Figure \ref{sim_img}) with observations (Figure \ref{obs_img}), we see that the true emission source in the 12 July 2017 event was located not far from the disk center ($\theta_0 \lesssim 10^{\circ}$) and was probably associated with the large active region (AR 12665). In this case, both the simulated source location and size agree well with the observed ones.

\subsection{Frequency dependence of the burst parameters}
\begin{figure}
\centerline{\includegraphics[width=0.97\linewidth]{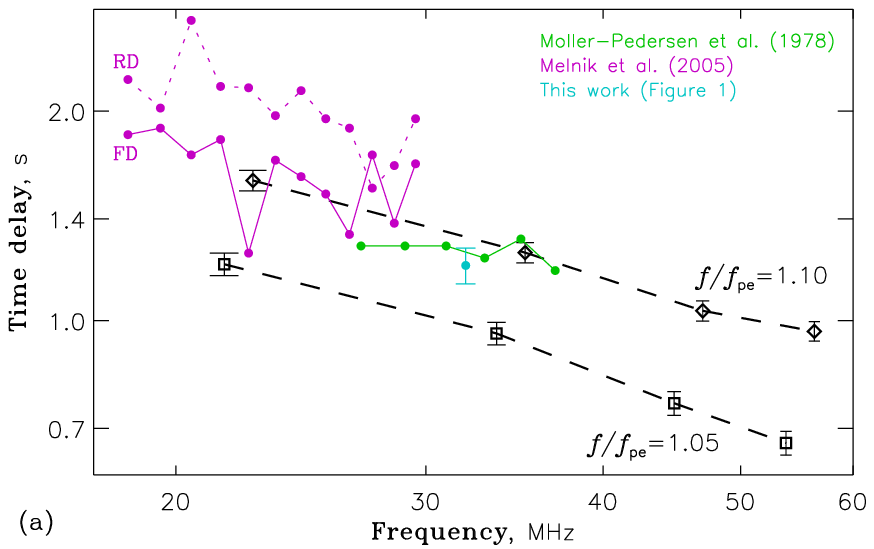}}
\centerline{\includegraphics[width=0.97\linewidth]{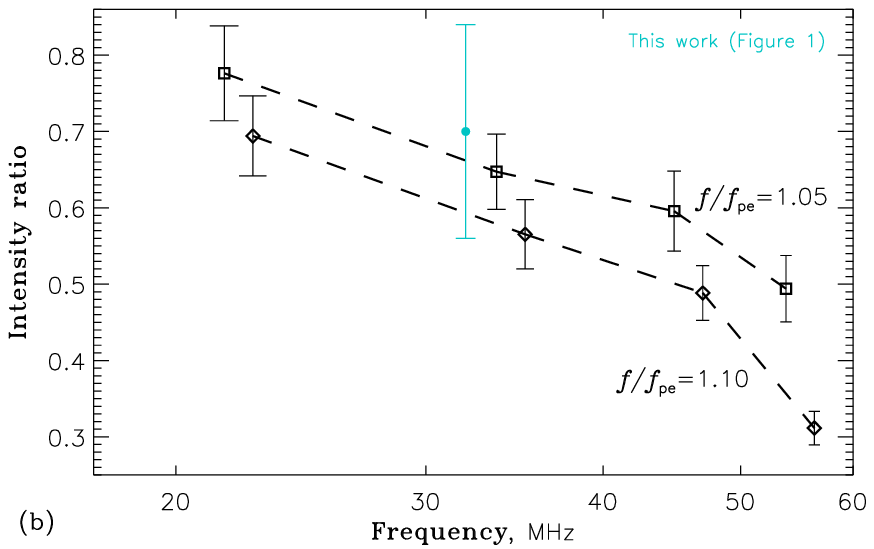}}
\caption{Parameters of drift-pair bursts vs. emission frequency. a) Time delay between the burst components. b) Relative intensity of the second component. Black lines show the simulation results for an emission source located at longitude $\theta_0 = 10^{\circ}$, $\epsilon=0.8$, $\alpha=0.1$, and initial emission to plasma frequency ratios $f/f_{\mathrm{pe}}(r_0)=1.05$ and $1.10$ (marked by $\Box$ and $\Diamond$ symbols, respectively). The observations are plotted as indicated by the legend. Error bars represent one standard deviation.}
\label{freq_dependence}
\end{figure}

Similar simulations were performed for several emission frequencies between 20 and 60 MHz and different values of $f/f_{\mathrm{pe}}(r_0)$. For brevity, we do not show individual images and time profiles, but summarize the obtained trends and present them in Figure \ref{freq_dependence} together with the available observational data from \citet{moller_1978} and \citet{melnik_2005}\footnote{\citet{moller_1978} and \citet{melnik_2005} present only the time delays between the burst components, and do not report the error bars of these values. \citet{melnik_2005} report the time delays separately for the bursts with forward (negative) and reverse (positive) frequency drifts.}, and with the observed characteristics of the drift-pair burst shown in Figure \ref{obs_prof}.

Figure \ref{freq_dependence}a presents the time delay between the burst components computed for several frequencies in the $20-60$ MHz range for initial ratios of the emission frequency to the local plasma frequency $f/f_{\mathrm{pe}}(r_0)=1.05$ and $1.10$, $\epsilon=0.8$, $\alpha=0.1$, and a source located at $\theta_0 = 10^{\circ}$. The delay is shorter for lower values of $f/f_{\mathrm{pe}}(r_0)$, because in this case the source is located closer to the radio-wave reflection surface (where the emission frequency approaches the local plasma frequency), and therefore the additional path traveled by the reflected signal (relative to the direct one) is shorter. This dependence can potentially be used as a diagnostic tool: by measuring the time delay between the components of drift-pair bursts, we can estimate the relative frequency $f/f_{\mathrm{pe}}$ in their emission sources and hence find the characteristic wave number $k$ of the Langmuir waves responsible for the plasma emission.

The delay between the components gradually decreases with an increase in the emission frequency (approximately $\propto f^{-1/2}$), because for a fixed $f/f_{\mathrm{pe}}(r_0)$ ratio the higher-frequency sources are located closer to the reflection surface, too. Currently, the observational data on the frequency dependence of the component delay in drift-pair bursts are ambiguous, e.g., \citet{delaNoe_1971} and \citet{moller_1978} concluded that the delay is independent of the frequency. On the other hand, \citet{melnik_2005} analyzed a number of drift-pair bursts in the frequency range of $18-30$ MHz and found a slow decrease in the time delay with frequency. Notably, both the delay values and their dependence on frequency reported by \citet{melnik_2005} agree well with the prediction of our model (see Figure \ref{freq_dependence}a) for an $f/f_{\mathrm{pe}}(r_0)$ ratio of about $1.10$, or slightly higher. Varying the anisotropy level $\alpha$ has almost no effect on the delay between the components.

Figure \ref{freq_dependence}b presents another important characteristic of the drift-pair bursts: the intensity ratio of the second (reflected) and first (direct) components, or the relative intensity of the second component; the simulation parameters are the same as in Figure \ref{freq_dependence}a. As discussed above, the formation of drift-pair bursts with components of comparable intensities requires strong anisotropy of the plasma turbulence. In addition, as Figure \ref{freq_dependence}b indicates, the relative intensity of the second component decreases with increasingemission frequency, which is caused by an increasing collisional damping: the reflected signal, which travels a longer distance and in a denser plasma, experiences a stronger free-free absorption than the direct signal. This effect helps explain why the drift-pair bursts are observed predominantly at low frequencies (below $\sim 100$ MHz): at higher frequencies, the collisional absorption becomes so strong that the reflected component can no longer be resolved. Varying the initial emission to plasma frequency ratio $f/f_{\mathrm{pe}}(r_0)$ has a relatively weak effect: in the considered case of strong anisotropy, a slightly higher relative amplitude of the second burst component is achieved for lower values of $f/f_{\mathrm{pe}}(r_0)$.

\section{Discussion and summary}\label{discussion}
We applied the newly-developed Monte Carlo technique \citep[see][]{kontar_2019} to perform large-scale simulations of radio-wave propagation in the anisotropic turbulent solar atmosphere, and compared them with observations of drift-pair bursts. The simulations have demonstrated that the used model can quantitatively reproduce the observed features and key properties of drift-pair bursts.

The key feature of drift-pair bursts is that their components are repeated in time rather than shifted in frequency. This feature immediately prompted an explanation that the second component is an echo caused by reflection of radio waves from lower layers of the solar atmosphere \citep{roberts_1958}. However, to produce the observed delays ($\sim 1-2$ s) between the burst components, the emission source had to be located at a rather large distance from the reflection layer (which implied harmonic plasma emission mechanism). As a result, the early reflection-based models predicted that a) the reflected signal should be weaker and more diffuse than the direct one, b) apparent source positions of the first and second components should be considerably different, and c) the delay between the components should increase with the emission frequency. These conclusions were not supported by observations.

We highlight that the first models of drift-pair bursts a) neglected \citep[or only partially considered, see][]{riddle_1974} radio-wave scattering, and b) assumed that the emission patterns are isotropic (see \citealt{mclean_1985} for a review). However, strong scattering of radio waves in the metric wavelength range is able to significantly affect both the time profiles (causing an additional delay, due to the fact that the radio waves no longer propagate along straight paths) and the apparent source positions \citep[see][]{kontar_2017, kontar_2019}. Moreover, anisotropic scattering in combination with large-scale refraction can provide a strong emission directivity even if the emission was originally (in the source) isotropic. Notably, the scattering would affect both the direct and reflected radio signals.

Our simulations have demonstrated that the combined effect of refraction/reflection and anisotropic scattering can indeed result in the formation of nearly exact echoes of radio signals, although only under certain conditions. Additional delay caused by scattering is able to provide the observed delays between the burst components even for the fundamental plasma emission ($f/f_{\mathrm{pe}}\simeq 1.10$) and relatively small distances between the source and the reflection layer. The latter effect (together with the fact that both the direct and reflected rays experience the same scattering and the observed emission originates from the last-scattering surface) results in coinciding apparent source positions of the burst components. The delays between the burst components slowly decrease with emission frequency, which qualitatively and quantitatively agrees with the observations by \citet{melnik_2005} and \citet{kuznetsov_2019}. The most important factor affecting the delay between the burst components is the ratio of the emission to plasma frequency in the emission source $f/f_{\mathrm{pe}}(r_0)$, which determines the path difference between the direct and reflected signals; therefore observations of drift-pair bursts can be used to diagnose the plasma emission mechanism. Evidently, the repetitive structure can only be observed if the bursts themselves are short enough, specifically, much shorter than the delay between the direct and reflected signals.

Essentially, the formation of drift-pair bursts requires an anisotropic scattering which implies anisotropic plasma density fluctuations; we considered the case of preferable scattering in the direction perpendicular to the local magnetic field (i.e., to the local radial direction). While both random turbulence and deterministic ``fibrous'' structures have been proposed as the cause of anisotropic scattering, turbulence seems more likely because---under typical coronal conditions---it is able to provide a much higher scattering rate. For weak or moderate anisotropy, the reflected signal becomes diffused and weakened, in accordance with earlier estimations. In contrast, strong scattering anisotropy has a focusing effect and results in high directivity of the emission, which makes the direct and reflected signals very similar both in duration and in amplitude; the main factor resulting in attenuation of the reflected signal is collisional damping of the emission. The above-mentioned requirements allow us to explain why other types of solar radio bursts do not produce noticeable echo components at these frequencies \citep[a problem first raised by][]{roberts_1958}: e.g., type III bursts are usually longer ($>1$ s), so that the echo component, if present, is lost in the tail of the direct component. On the other hand, the anisotropy should also be sufficiently high: an anisotropy level $\alpha>0.2$ results in a diffuse echo component with a low intensity (see Figure \ref{sim_e00}). Despite the high directivity of emission, drift-pair bursts (with nearly the same shapes of light curves and delays between the components) can be potentially formed for a wide range of source positions (both at the solar disk center and near the limb). However,

The formation of drift-pair bursts requires anisotropic density fluctuations with the typical wavelengths in the perpendicular direction to be an order of magnitude shorter than in the parallel direction. The directivity of escaping emission makes the sources located farther from the solar disk center to be fainter, because a lower fraction of the emission can reach the Earth. This affects both the intensity of radiation and, more importantly, the brightness of the fine spectral structures relative to the background continuum. Therefore, the drift-pair bursts are more likely to be detected when their sources are located near the disk center. This result naturally explains the known statistics of center-to-limb variation \citep{mollerpedersen_1974}.

The drift-pair bursts have been observed in a limited range of frequencies ($\sim 10-100$ MHz). The low-frequency boundary could be instrumental: the ionospheric cutoff affects ground-based observations, while space-based radio instruments tend to have lower time and frequency resolutions, and a lower sensitivity. On the other hand, the high-frequency boundary is likely due to collisional damping, which affects the reflected signal more strongly. Since the damping increases with the plasma frequency, the relative amplitude of the reflected component decreases accordingly, until (at $f\gtrsim 100$ MHz) the reflected signal becomes too faint to be distinguished.

The frequency drift rates of the drift-pair bursts are intermediate between those of type II and type III bursts at the same frequency. Thus, assuming the same plasma emission mechanism, the speed of an exciting agent should be an order of magnitude higher than the speed of magnetohydrodynamic shock waves exciting the type II bursts, but a few times lower than the speed of relativistic electrons producing the type III bursts \citep{melrose_1982}. A likely candidate are the whistler wave packets which have typical group speeds of about $(21-28)v_{\mathrm{A}}$, where $v_{\mathrm{A}}$ is the Alfv\'en speed \citep{kuijpers_1975}. E.g., for the burst shown in Figure \ref{obs_prof}, we estimate the drift-pair exciter speed as $v\simeq (\partial f/\partial t)/(\partial f_{\mathrm{pe}}/\partial r)\simeq 20\,000$ km $\textrm{s}^{-1}$ at $f\simeq f_{\mathrm{pe}}\simeq 30$ MHz. This speed (assuming the whistler explanation) requires a magnetic field of about 1 G, which seems quite achievable in the drift-pair burst sources. The whistler packets can propagate both upwards and downwards, producing bursts with negative and positive frequency drifts, respectively, sometimes simultaneously in the same event. This implies that the whistler packets are generated immediately within the source region of drift-pair bursts. On the other hand, the particular emission mechanism (i.e., how whistler packets can produce or modulate the radio emission) and relation of the drift-pair bursts to the accompanying type III storms, require further investigation.

\acknowledgments
A.A.K. acknowledges partial support from budgetary funding of Basic Research program II.16 and the program KP19-270 of the RAS Presidium.
N.C. was supported by the STFC grant ST/N504075/1.
E.P.K. was supported by the STFC grants ST/P000533/1 and ST/T000422/1.
The authors acknowledge the support by the international team grant (\url{http://www.issibern.ch/teams/lofar/}) from ISSI Bern, Switzerland.
This paper is based (in part) on data obtained from facilities of the International LOFAR Telescope (ILT) under project code LC8\_027.
LOFAR \citep{vanHaarlem_2013} is the Low-Frequency Array designed and constructed by ASTRON.

\bibliographystyle{aasjournal}
\bibliography{references}
\end{document}